\documentclass[fleqn,10pt]{wlscirep}

\usepackage[justification=justified, font=footnotesize]{caption}
\usepackage{graphicx}
\usepackage{floatrow}
\usepackage[labelformat=simple]{subfig}
\usepackage[labelformat=empty]{caption}
\usepackage{hyperref}

\title{Cu$_2$O Microcrystals Grown on Silicon as Platform for Quantum-Degenerate Excitons and Rydberg States}

\author[1,*]{Stephan Steinhauer}
\author[1]{Marijn A. M. Versteegh}
\author[2]{André Mysyrowicz}
\author[3]{Birgit Kunert}
\author[1]{Val Zwiller}
\affil[1]{Department of Applied Physics, KTH Royal Institute of Technology, SE-100 44 Stockholm, Sweden}
\affil[2]{Laboratoire d’Optique Appliquée, ENSTA ParisTech, CNRS, Ecole Polytechnique, F-91762 Palaiseau, France}
\affil[3]{Institute of Solid State Physics, Graz University of Technology, A-8010 Graz, Austria}

\affil[*]{e-mail: \href{mailto:ssteinh@kth.se}{ssteinh@kth.se}}

\begin{abstract}
\textbf{
Cuprous oxide (Cu$_2$O) is a semiconductor with large exciton binding energy and significant technological importance in applications such as photovoltaics and solar water splitting. It is also a superior material system for quantum optics that enabled the observation of two intriguing phenomena, i.e. Rydberg excitons as solid-state analogue to highly-excited atomic states and dense exciton gases showing quantum degeneracy when approaching the phase transition to Bose-Einstein condensation. Previous experiments focused on natural bulk crystals due to major difficulties in growing high-quality synthetic samples. Here, we present Cu$_2$O microcrystals with excellent optical material quality capable of hosting both quantum-degenerate excitons and excited Rydberg states. Growth of Cu$_2$O with exceedingly low point defect levels was achieved on silicon by a scalable thermal oxidation process compatible with lithographic patterning. Using the latter, we demonstrate Rydberg excitons in site-controlled Cu$_2$O microstructures, paving the way for a plethora of applications in integrated quantum photonics.
}
\end{abstract}

\begin{document}
\flushbottom
\maketitle
\thispagestyle{empty}

Light-matter interactions in the direct-band-gap semiconductor cuprous oxide (Cu$_2$O) are widely governed by excitons, quasi-particles arising from electron-hole Coulomb interactions, which can be observed up to room temperature due to their large binding energy. As a result of the unique excitonic properties, intriguing condensed-matter quantum phenomena have been demonstrated, e.g. quantum-degenerate exciton gases\cite{Snoke1987,Snoke1990BEC} and giant Rydberg excitons\cite{Kazimierczuk2014} exhibiting signatures of quantum coherences \cite{Gruenwald2016} and quantum chaotic behavior \cite{Assmann2016}. As the exciton ground state - the so-called paraexciton - is decoupled from the radiation field in unstrained Cu$_2$O, paraexcitons can reach lifetimes up to microseconds rendering them a prime candidate for excitonic Bose-Einstein condensation \cite{Snoke2014}. Apart from quantum optics, recent reports have reinforced the significant potential of Cu$_2$O as a low-cost, non-toxic material in areas such as photocatalysis \cite{Wu2017}, solar water splitting \cite{Pan2018} and photovoltaic devices \cite{Minami2016}. Various methods have been reported for the growth of Cu$_2$O thin films and single crystals \cite{Meyer2012}, in particular molecular beam epitaxy \cite{Li2013,Kracht2016}, vapor phase transport \cite{Brittman2014}, magnetron sputtering \cite{Yin2005} in combination with thermal annealing \cite{Bergum2018}, thermal oxidation \cite{Mani2009} and the floating zone method \cite{Chang2013}. However, state-of-the-art quantum optics experiments still focus on natural bulk crystals originating from mines, clearly underlining that significant progress in Cu$_2$O growth is required to surpass inherent limitations of natural samples. Recent literature emphasized that for the observation of Rydberg states with principal quantum numbers higher than the current record value of \textit{n}=25 samples with lower impurity concentrations are required \cite{Heckotter2018}. In addition, scalable fabrication techniques suitable for obtaining micro- / nanostructures and compatible with standard silicon processing are needed to deploy the full potential of this material in advanced device technologies, for instance integrated quantum photonic circuits.

In this work, we present Cu$_2$O growth on silicon by a scalable thermal oxidation process, which resulted in high-quality microcrystal structures with exceedingly low point defect and impurity levels. We achieve a dense exciton gas in single Cu$_2$O microcrystals at milli-Kelvin temperatures under continuous-wave excitation, showing quantum degeneracy close to the phase boundary to Bose-Einstein condensation. Furthermore, we demonstrate luminescence from excited \textit{np} Rydberg states in lithographically site-controlled structures, exhibiting excellent agreement with a hydrogen-like quantum number dependence. Due to the exceptional optical material quality and the unique excitonic properties, our Cu$_2$O microcrystals are highly promising for enabling new photonic device architectures relevant for quantum information processing and quantum sensing.

\subsection*{Growth and structural/photoluminescence properties}
\begin{figure}
  \centering
  \sidesubfloat[]{\includegraphics[width=0.9\textwidth]{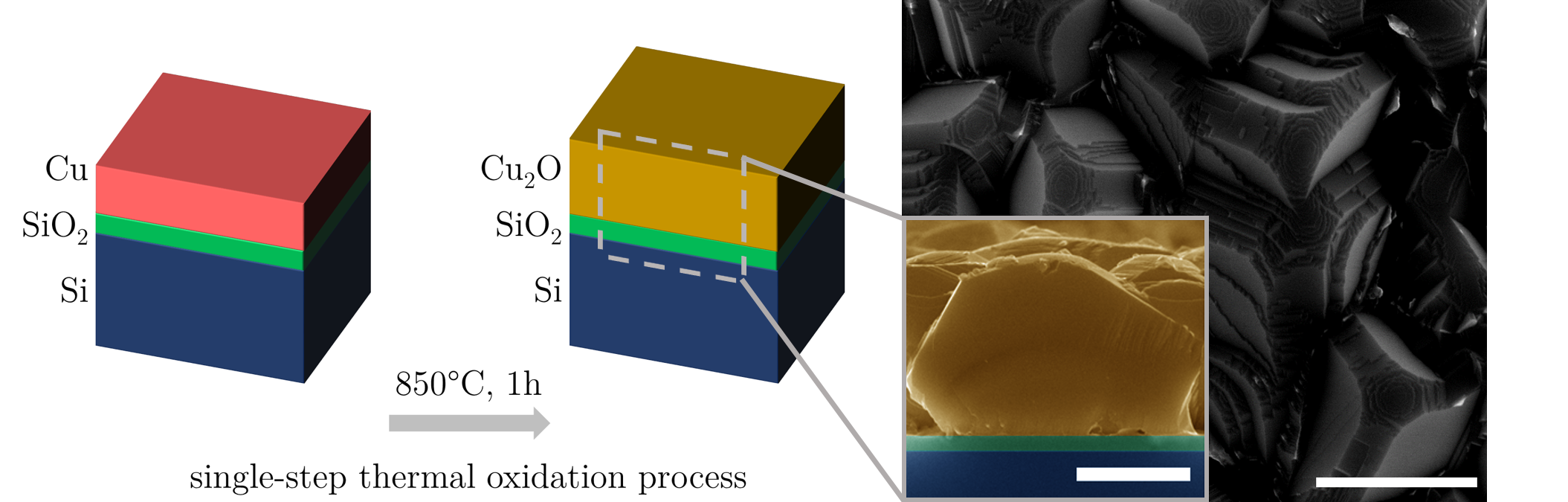}\label{fig1:sub1}}%
  \vspace{8mm}
  \sidesubfloat[]{\includegraphics[width=0.45\textwidth]{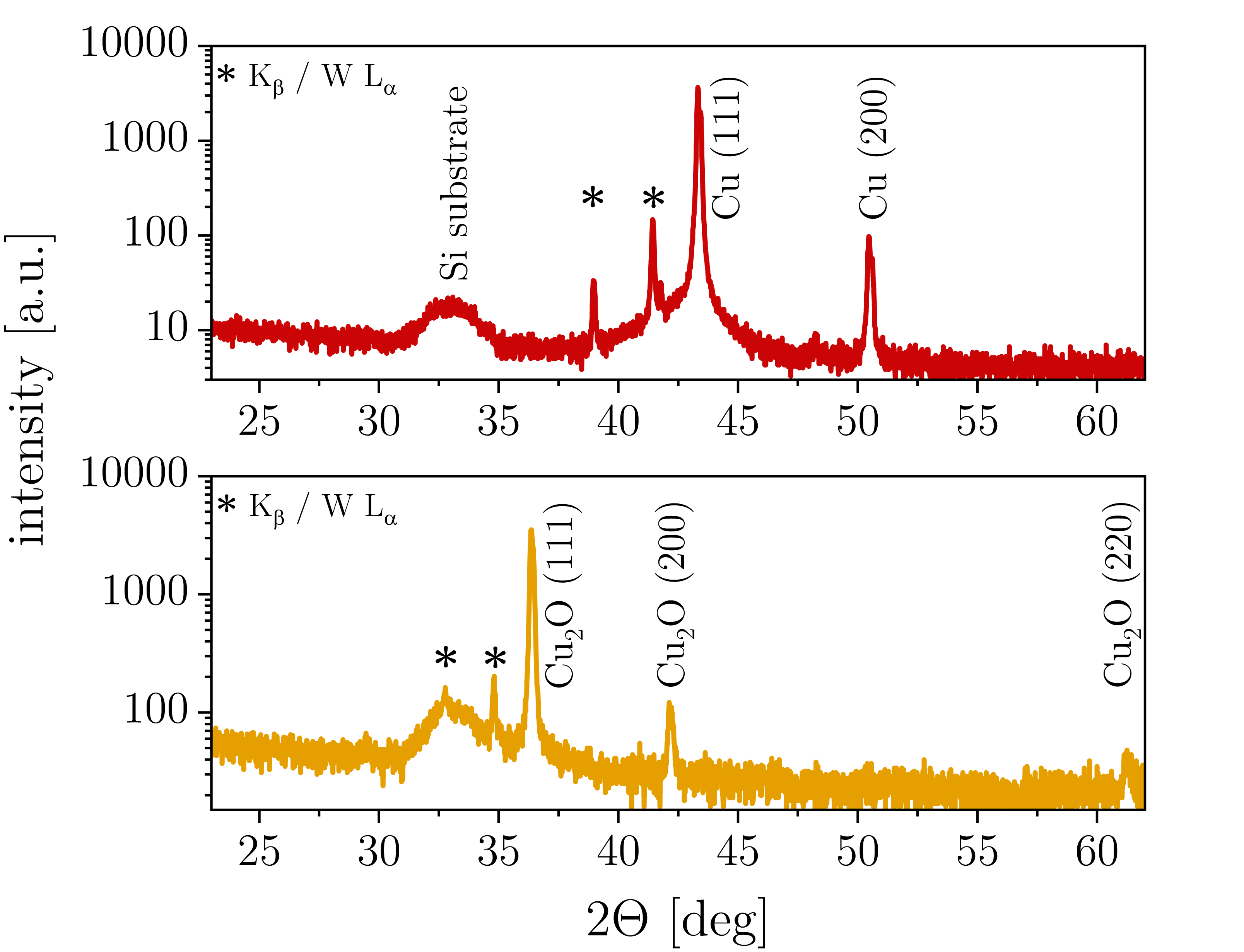}\label{fig1:sub2}}%
  \sidesubfloat[]{\includegraphics[width=0.45\textwidth]{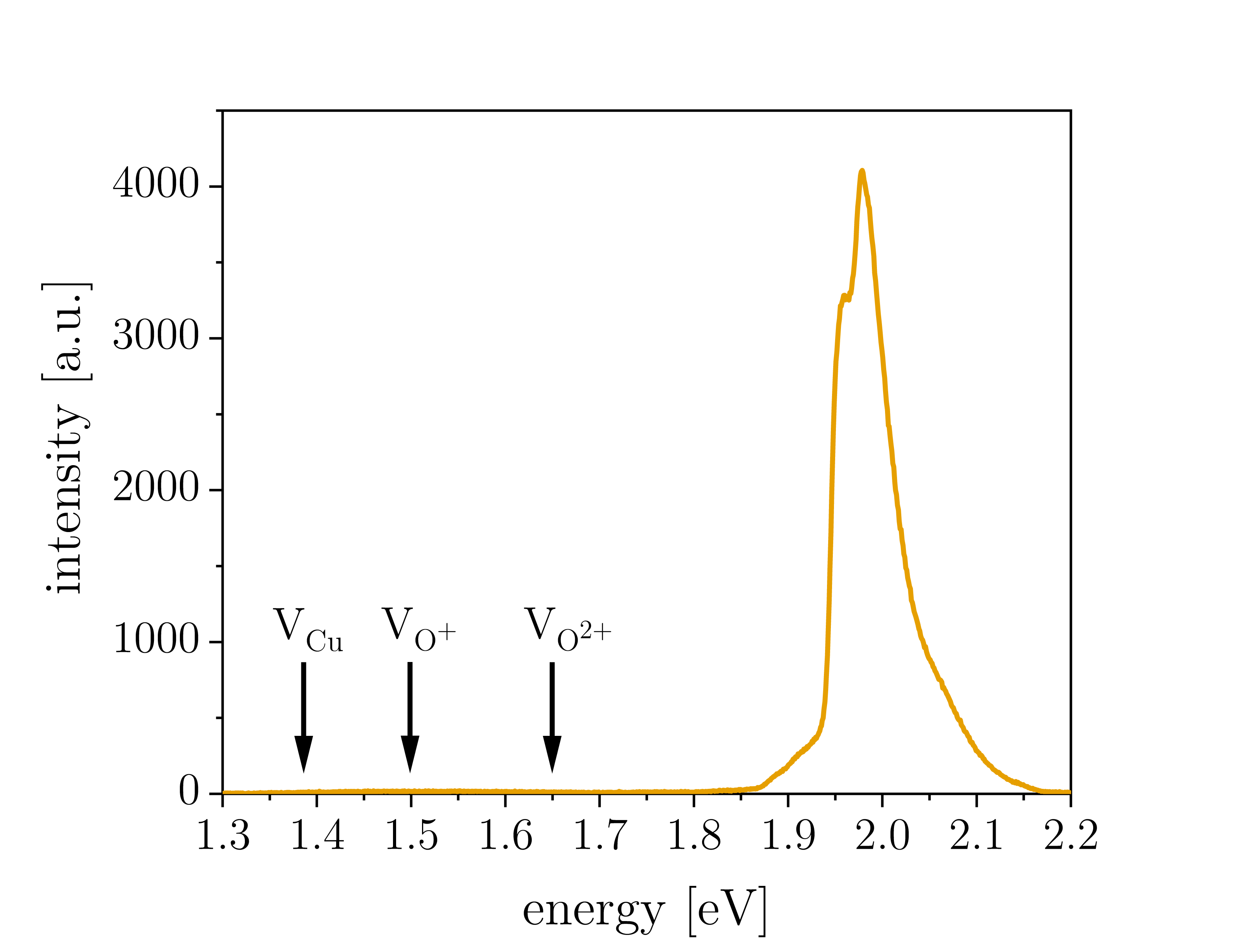}\label{fig1:sub3}}%
  \caption{\textbf{Fig.1 \textbar Growth and characterization of Cu$_2$O microcrystals. a}, Schematics of sample layer structure before and after the thermal oxidation process (left). Top-view scanning electron micrograph of Cu$_2$O microcrystals after thermal oxidation at 850$^\circ$C for 1\,h at pressures of 1\,mbar synthetic air (right); the bottom inset shows a corresponding cross-sectional image (scale bars 1\,$\mu$m). \textbf{b}, X-ray diffraction of as-deposited copper film (top) and Cu$_2$O film after thermal oxidation (bottom). \textbf{c}, Room-temperature photoluminescence spectroscopy of Cu$_2$O microcrystal under continuous-wave laser excitation (532\,nm). The emission around 2\,eV is due to the recombination of free excitons.}
  \label{fig1}
\end{figure}

The growth of Cu$_2$O microcrystals relied on a scalable single-step thermal oxidation process schematically depicted in Fig.\ref{fig1:sub1}. Copper films (thickness $\sim$700\,nm) were deposited by electron beam evaporation on silicon substrates covered with a thermal SiO$_2$ layer. Thermal oxidation in a tube furnace resulted in Cu$_2$O films with microcrystalline morphology, which can be seen in the top-view and cross-sectional scanning electron microscopy images. The Cu$_2$O microcrystals showed terrace-like structures on the surface (Supplementary Fig.S1) and faceted grains with sizes in the $\mu$m range. As-deposited copper and samples after thermal oxidation to Cu$_2$O were characterized by X-ray diffraction (XRD) measurements (Fig.\ref{fig1:sub2}) to determine their phase composition. For as-deposited copper, the expected face-centered cubic structure and texturing along the [111] direction was found. The presented single-step thermal oxidation method resulted in phase-pure Cu$_2$O with cubic cuprite structure. For samples fabricated with different growth conditions comparable microcrystal morphology as well as similar XRD results were obtained (Supplementary Fig.S2).
\floatsetup[figure]{style=plain,subcapbesideposition=top}
Thermal oxidation of copper can lead to different oxide phases \cite{Meyer2012}; Cu$_2$O growth was reported to proceed via the random nucleation of islands, which exhibit a cube-on-cube epitaxial relationship at the metal-oxide interface (low oxygen partial pressures) \cite{Zhou2013} or non-epitaxial orientations (above critical oxygen partial pressures depending on the copper surface) \cite{Luo2012}. For the range of experimental parameters corresponding to the growth conditions used in this study (800-850$^\circ$C, p$\sim$1\,mbar of synthetic air) the initial stages of copper oxidation include epitaxial oxide formation accompanied by rapid two-dimensional growth \cite{Gattinoni2015}. The XRD results before and after thermal oxidation (Fig.\ref{fig1:sub2}) show pronounced texturing in \{111\} direction in both cases, indicating that the oxidation proceeds via an epitaxial relationship of Cu$_2$O \{111\} $\parallel$ Cu \{111\}. This relationship is in accordance with literature reports on the thermal oxidation of copper thin films \cite{Zhou2013} and nanoparticles with sizes down to few nanometers \cite{LaGrow2017}. Similar to a previous report on epitaxial Cu$_2$O growth on MgO \cite{Yin2005}, the observed terrace-like structures on the Cu$_2$O surfaces are suggesting a two-dimensional growth mode for individual microcrystals.

The optical material quality of the Cu$_2$O microcrystals was initially assessed by means of room-temperature photoluminescence spectroscopy. Distinct free exciton emission was observed (Fig.\ref{fig1:sub3}), exhibiting a characteristic lineshape resulting from multiple phonon-assisted recombination processes with spectral overlap \cite{Li2013}. At photon energies attributed to copper vacancies or single/double-charged oxygen vacancies \cite{Meyer2012} no marked luminescence was observed. Additional data for microcrystalline Cu$_2$O films from different batches and grown under different conditions can be found in Supplementary Fig.S3, which shows spectra with very similar characteristics. Hence, the presented single-step thermal oxidation process is a robust method for the realization of microcrystalline Cu$_2$O films with excellent optical material quality. 

\subsection*{Point defects, bound excitons and yellow 1\textit{\textbf{s}} excitons}
\begin{figure}[h]
  \centering
  \sidesubfloat[]{\includegraphics[width=0.45\textwidth]{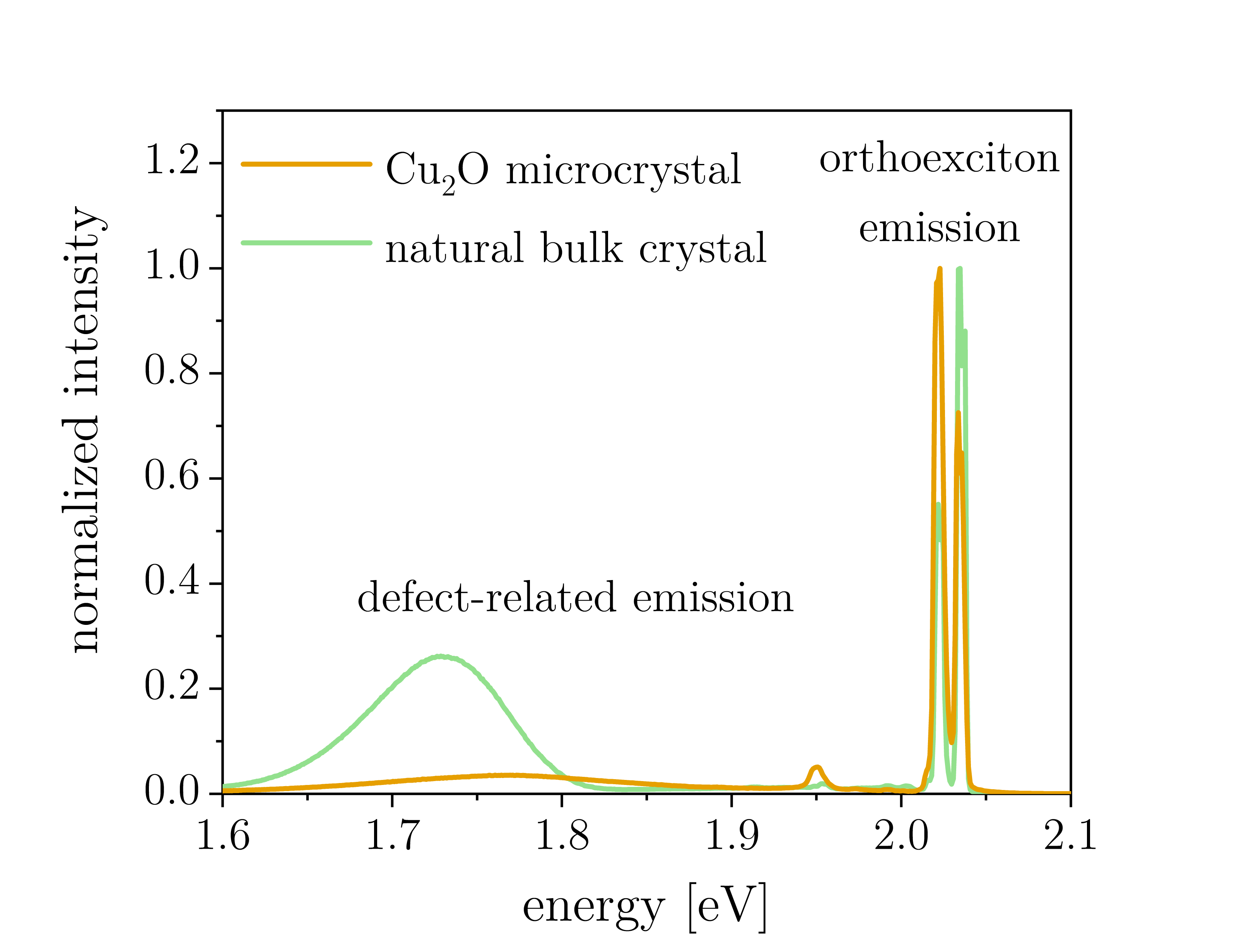}\label{fig2:sub1}}%
  \sidesubfloat[]{\includegraphics[width=0.45\textwidth]{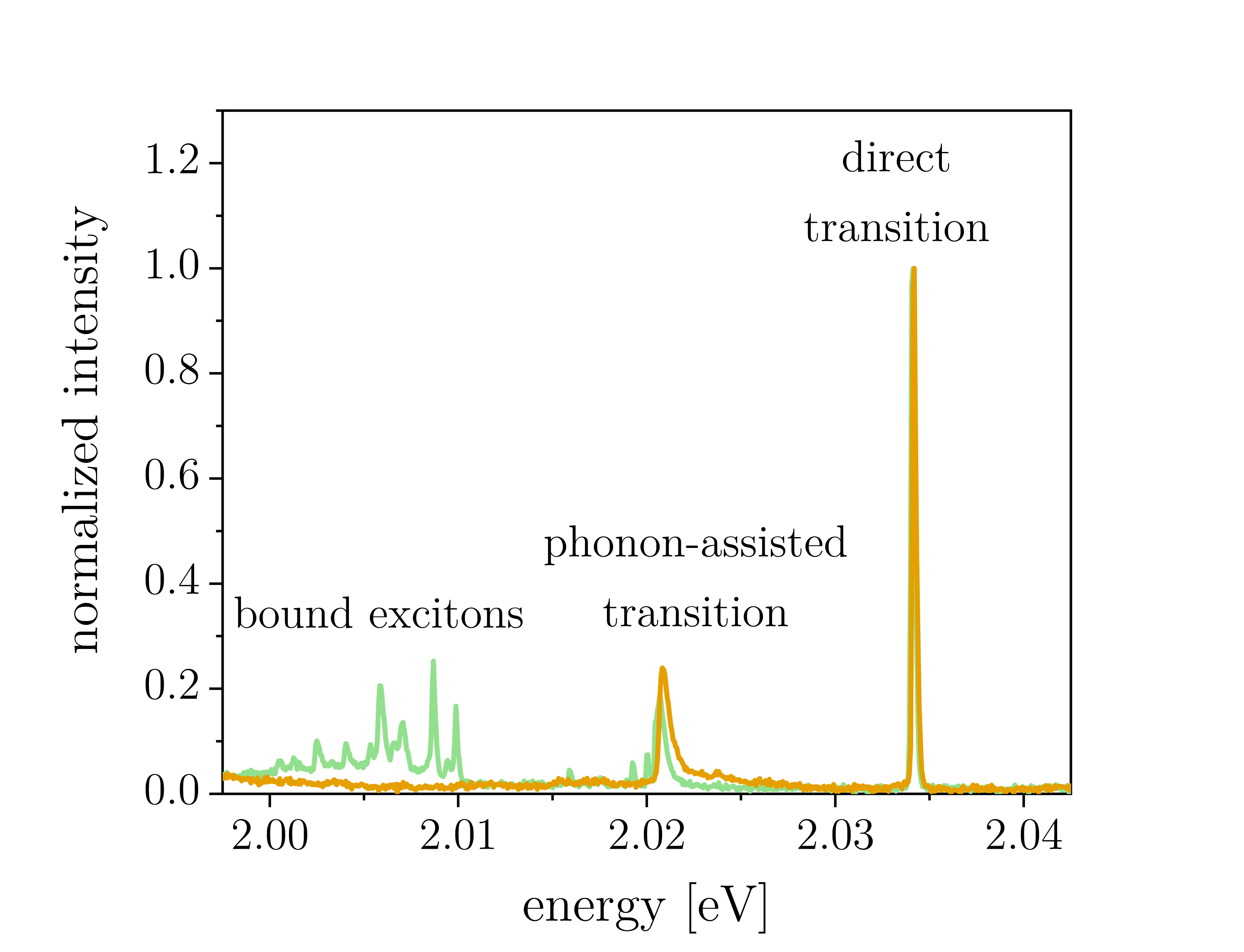}\label{fig2:sub2}}%
  \vspace{1mm}
  \sidesubfloat[]{\includegraphics[width=0.45\textwidth]{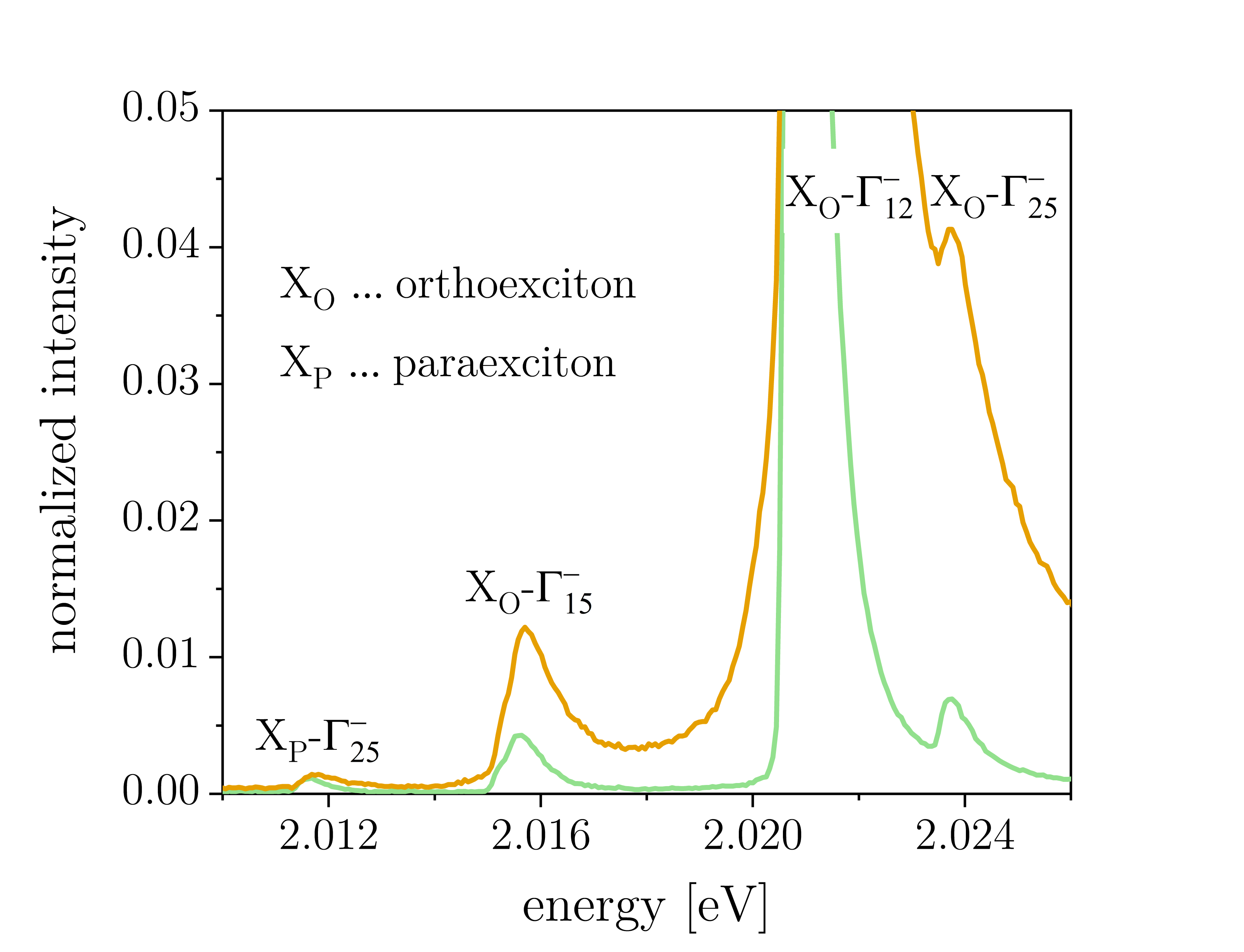}\label{fig2:sub3}}%
  \sidesubfloat[]{\includegraphics[width=0.45\textwidth]{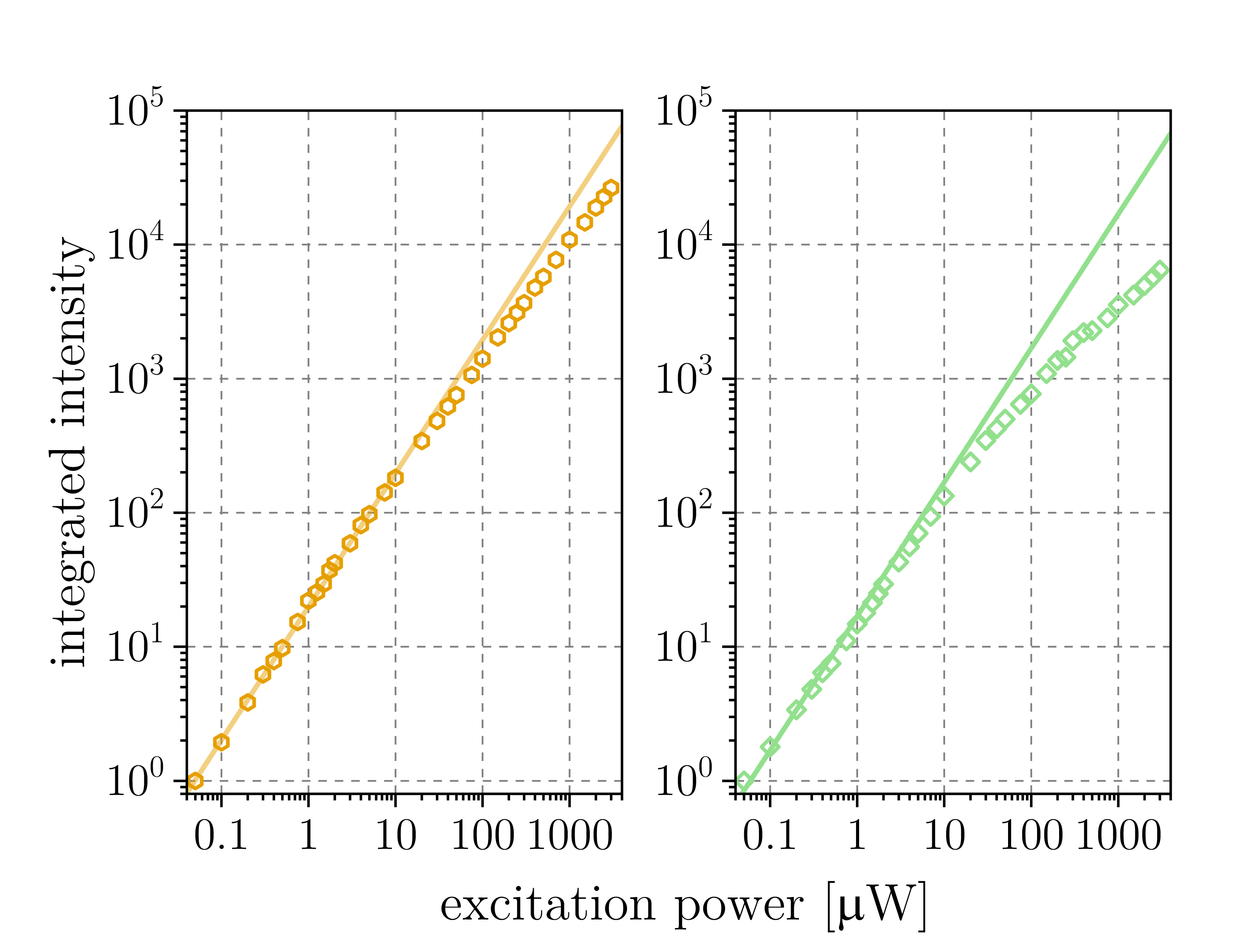}\label{fig2:sub4}}%
  \caption{\textbf{Fig.2 \textbar Photoluminescence spectroscopy of Cu$_2$O microcrystals at milli-Kelvin temperatures benchmarked to measurements on a natural bulk crystal. a}, Normalized photoluminescence of excitons and point defects (excitation power 50\,$\mu$W). \textbf{b}, Emission from different excitonic transitions and bound excitons at low excitation power of 50\,nW (spectra normalized to direct transition). \textbf{c}, Phonon-assisted transitions of ortho- and paraexcitons (spectra normalized to direct transition; excitation power 50\,$\mu$W). \textbf{d}, Excitation power dependence of integrated emission in an energy range covering bound excitons, phonon-assisted transitions and the direct quadrupole line. Solid lines are a linear fit (Cu$_2$O microcrystal, left) and a linear curve as guide to the eye (natural bulk crystal, right), respectively.}
  \label{fig2}
\end{figure}

Local deviations from the ideal cuprite crystalline structure, e.g. vacancy point defects, extrinsic impurities or microscopic strain, have a significant impact on the relaxation of excitons in Cu$_2$O and the related photon emission. Photoluminescence spectroscopy experiments were conducted in a dilution refrigerator (sample stage base temperature around 40\,mK) to assess the optical material quality of Cu$_2$O microcrystals at milli-Kelvin temperatures. The results were compared to natural bulk Cu$_2$O as benchmark (crystal originates from a geological sample used in previous literature \cite{Mysyrowicz1979}). Spectra normalized to the yellow 1\textit{\textbf{s}} orthoexciton emission that were acquired at a laser power of 50\,$\mu$W (corresponding to a peak intensity of 3\,kW/cm$^2$) and an excitation wavelength of 532\,nm are presented in Fig.\ref{fig2:sub1}, showing considerably reduced emission related to oxygen vacancies for the case of Cu$_2$O microcrystals. An emission feature around 1.95\,eV was observed for both Cu$_2$O microcrystals and the natural bulk sample, which has been reported repeatedly in literature. It was attributed to phonon-assisted transitions and defect emission in close spectral proximity \cite{Kracht2016,Frazer2017,Takahata2018} with the latter potentially being correlated with local strain in the sample \cite{Frazer2015}. Additional data for different samples and natural bulk crystal positions can be found in Supplementary Fig.S4. Low point defect densities are highly important for efficient cooling of the exciton gas in Cu$_2$O as trapping at defects is a limiting factor for exciton lifetimes. The latter were found to be significantly shortened for increasing oxygen vacancy concentrations \cite{Koirala2013}. Furthermore, it has been suggested that relaxation processes involving vacancies can lead to heating due to phonon emission \cite{Frazer2017}, which is detrimental for achieving cold exciton gas temperatures. In addition to intrinsic point defects, excitons and their luminescence properties may be influenced by the presence of extrinsic impurities. The latter can lead to the formation of bound excitons, which show multiple emission lines in the energy range around $\sim$2.00\,eV, below the phonon-assisted orthoexciton transition \cite{Jang2006}. In Fig.\ref{fig2:sub2} we directly compare photoluminescence spectra obtained from synthetic Cu$_2$O microcrystals and from the natural bulk crystal under identical experimental conditions (excitation power 50\,nW). It is evident that all peak-like features related to excitons bound to extrinsic impurities are absent in Cu$_2$O microcrystals, once more validating the excellent purity and material quality of our samples. Moreover, we assess the energy level structure of yellow 1\textit{\textbf{s}} excitons in Cu$_2$O microcrystals by monitoring luminescence from different phonon-assisted transitions (Fig.\ref{fig2:sub3}). The emission features were assigned according to previous literature \cite{Mysyrowicz1983} and the energy splitting of 1\textit{\textbf{s}} excitons into orthoexcitons and paraexcitons separated by 12\,meV in unstrained Cu$_2$O was validated. The influence of strain on the luminescence spectra of Cu$_2$O microcrystals is discussed in Supplementary Fig.S5. Exciton relaxation was further studied by assessing its excitation power dependence. For this purpose, the incident laser power was varied and the luminescence spectra were integrated in an energy range covering bound excitons, phonon-assisted transitions and the direct quadrupole line (Fig.\ref{fig2:sub4}). The integrated intensity of the Cu$_2$O microcrystal is in excellent agreement with a linear relationship (slope 0.995 $\pm$ 0.008) for excitation powers spanning over more than two orders of magnitude, showing slight sub-linear behaviour at elevated excitation levels. On the other hand, deviations from a linear power dependence were significantly more pronounced for the natural bulk crystal. Sub-linear power dependence of excitonic emissions in Cu$_2$O has been previously observed in experiments using natural bulk samples \cite{Ohara1999,Jang2006,Trauernicht1986,Stolz2012} and synthetic crystals grown by the floating zone method \cite{Karpinska2005} using various types of laser excitation. It has been attributed to an efficient non-radiative two-body recombination process, which can be explained by Auger decay \cite{Wolfe2014}, by the formation of short-lived biexcitons \cite{Jang2006b} or by exciton interconversion via spin exchange \cite{Kavoulakis2000}. The estimated recombination rates reported in literature vary considerably as the process is expected to be sensitive to Cu$_2$O crystal symmetry and the resulting band structure; hence it has been anticipated that Auger recombination is associated with broken band symmetries in the vicinity of impurities \cite{Snoke2014}, which would explain the less pronounced two-body decay in low-defect-density Cu$_2$O microcrystals. Hence, we consider the latter an ideal platform for studying high-density exciton gases and their quantum statistics, which will be detailed in the next section.

\floatsetup[figure]{style=plain,subcapbesideposition=top}

\newpage
\subsection*{Quantum-degeneracy of 1\textit{\textbf{s}} ortho- and paraexcitons}
Cu$_2$O has been considered the prime candidate for excitonic Bose-Einstein condensation in three-dimensional semiconductors due to several favourable properties, including large exciton binding energies of 150\,meV, small Bohr radii around 0.7\,nm leading to high Mott transition densities, as well as the same positive parity of the highest valence band and the lowest conduction band decoupling the exciton ground state (paraexciton) from photon interactions. Paraexcitons are thus particularly relevant for experiments related to Bose-Einstein condensation and have been predominantly studied using strain-induced confining potentials realized by the Hertzian stress technique of pressing a spherical object against a flat Cu$_2$O crystal surface \cite{Trauernicht1986,Yoshioka2011,Schwartz2012,Froehlich2018}. Decades of research on natural bulk samples have resulted in several reports of quantum degeneracy and Bose-Einstein condensation, which have been questioned by competing models assuming efficient two-body Auger recombination, inhomogeneous exciton distributions and exciton diffusion effects in macroscopic crystals \cite{Snoke2014}. We address this controversy by studying the exciton gas properties in Cu$_2$O microcrystals as a new experimental configuration providing confined geometries.

\begin{figure}[ht]
  \centering
  \sidesubfloat[]{\includegraphics[width=0.3\textwidth]{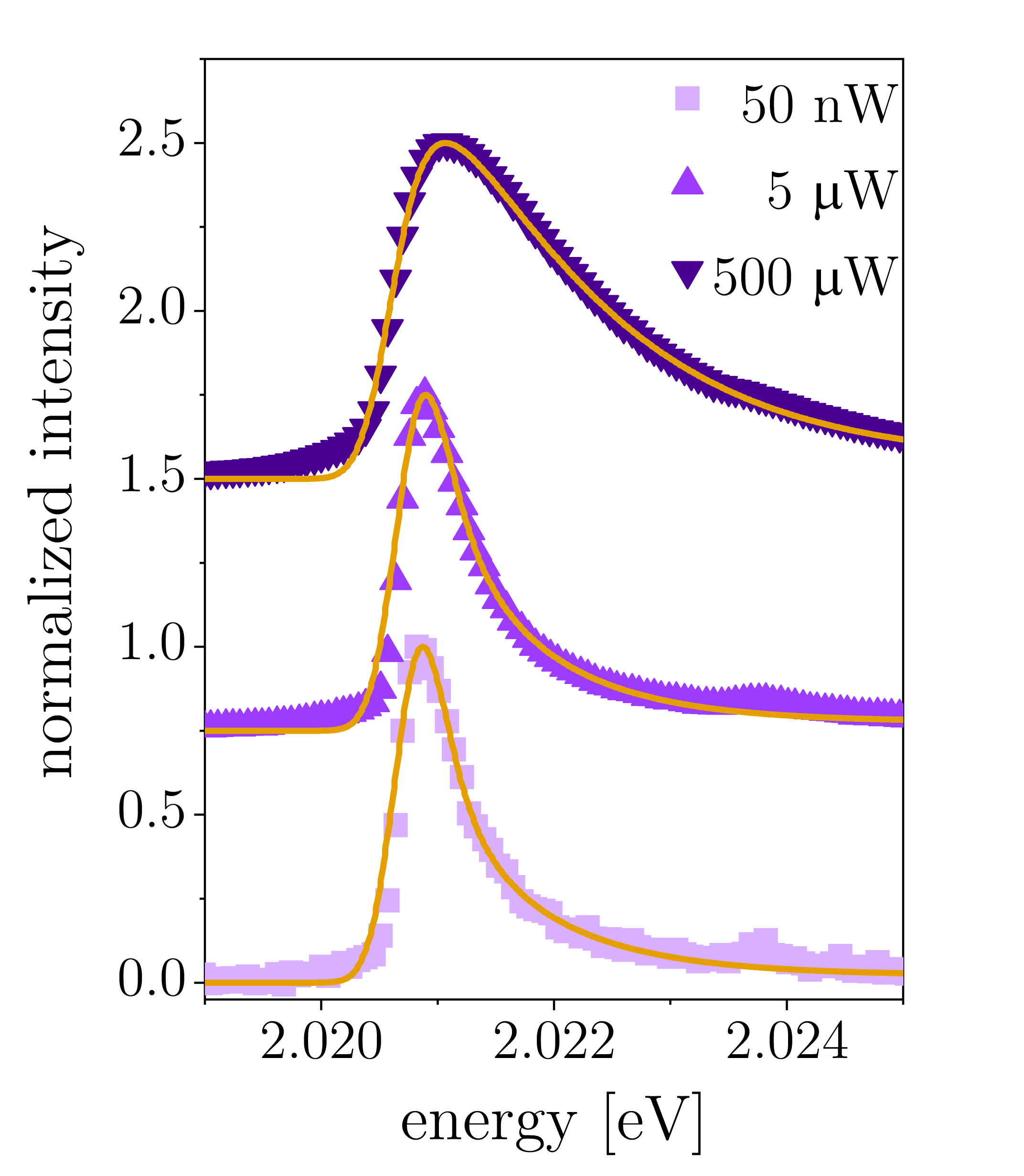}\label{fig3:sub1}}%
  \sidesubfloat[]{\includegraphics[width=0.3\textwidth]{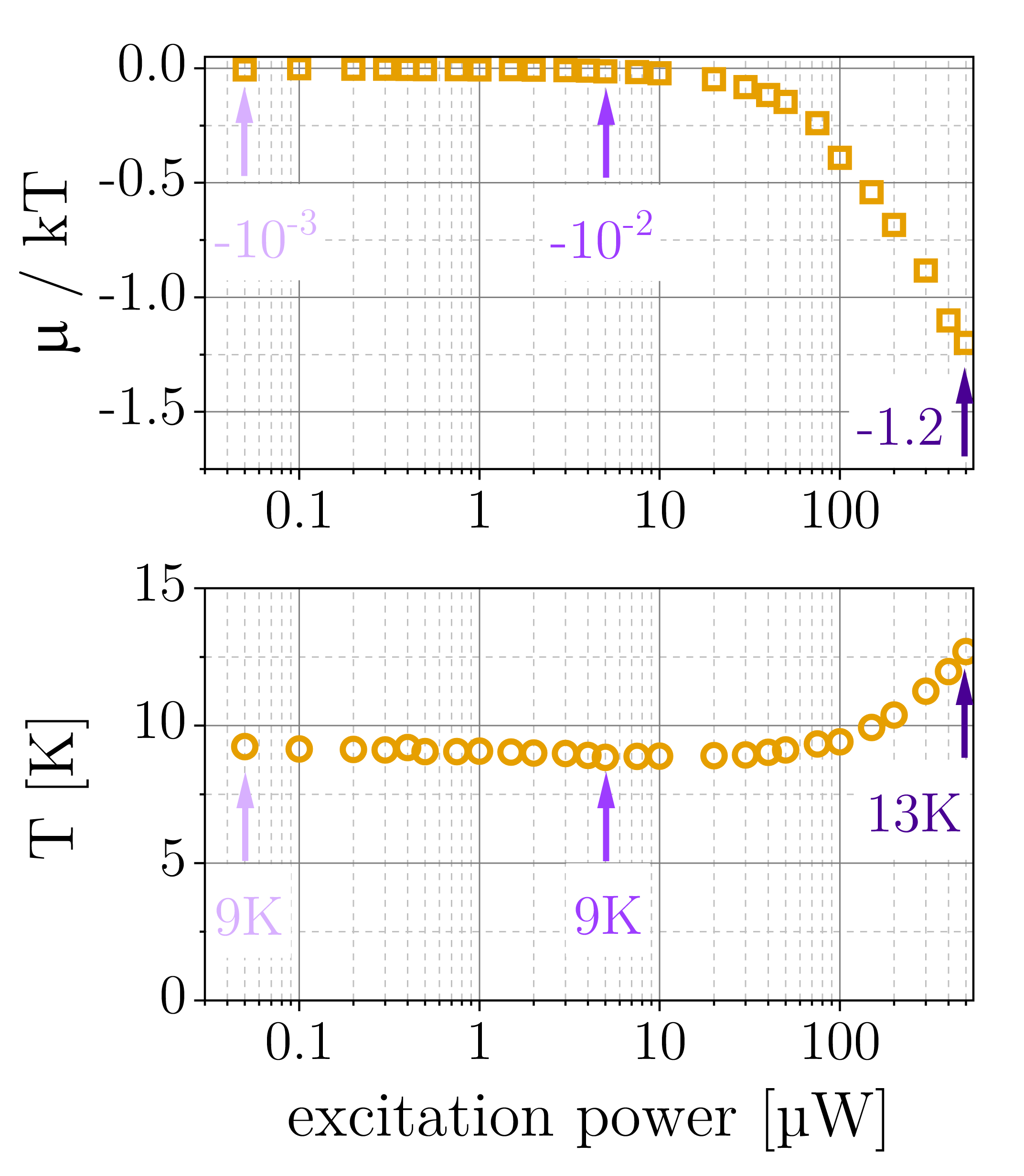}\label{fig3:sub2}}%
  \sidesubfloat[]{\includegraphics[width=0.3\textwidth]{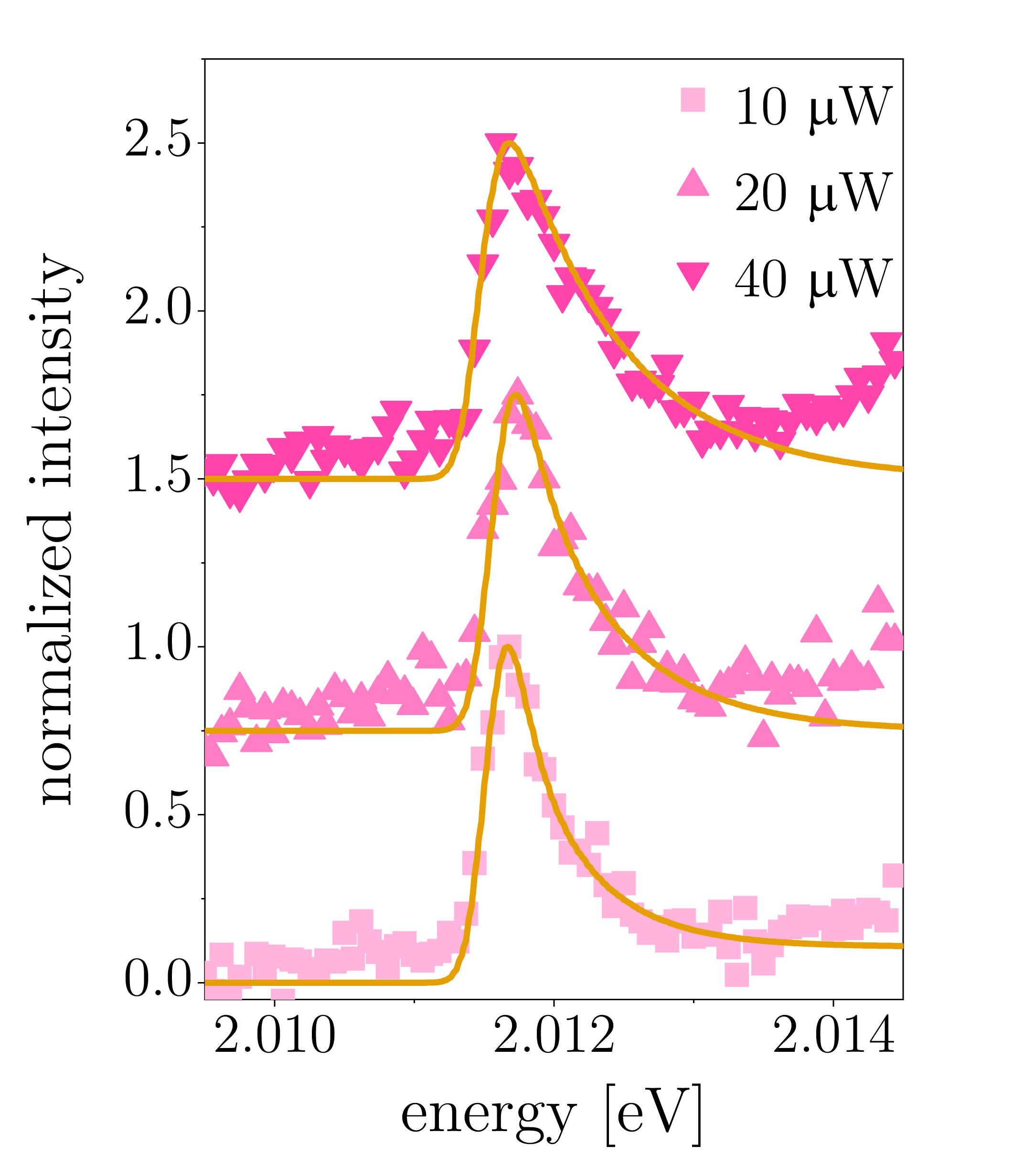}\label{fig3:sub3}}%
  \caption[\textbf{Fig.3 \textbar Quantum-degenerate exciton gas in Cu$_2$O microcrystals. a}, Photoluminescence spectra of phonon-assisted X$_O-\Gamma^-_{12}$ orthoexciton emission for three different excitation powers. The solid lines correspond to fits using a Bose-Einstein distribution function (spectra were offset vertically for clarity). \textbf{b}, Extracted fit parameters for the chemical potential $\mu$ and the temperature \textit{T} of the orthoexciton gas for excitation powers covering four orders of magnitude. The parameters for the spectra shown in a, are annotated in the graph. \textbf{c}, Phonon-assisted X$_P-\Gamma^-_{25}$ paraexciton transition for three different excitation powers with fits using a Bose-Einstein distribution function (spectra were offset vertically for clarity). Details related to the obtained fit parameters are described in the text.]{\textbf{Fig.3 \textbar Quantum-degenerate exciton gas in Cu$_2$O microcrystals. a}, Photoluminescence spectra of phonon-assisted X$_O-\Gamma^-_{12}$ orthoexciton emission for three different excitation powers. The solid lines correspond to fits using a Bose-Einstein distribution function (spectra were offset vertically for clarity). \linebreak \textbf{b}, Extracted fit parameters for the chemical potential $\mu$ and the temperature \textit{T} of the orthoexciton gas for excitation powers covering four orders of magnitude. The parameters for the spectra shown in a, are annotated in the graph. \textbf{c}, Phonon-assisted X$_P-\Gamma^-_{25}$ paraexciton transition for three different excitation powers with fits using a Bose-Einstein distribution function (spectra were offset vertically for clarity). Details on the obtained fit parameters are described in the text.}
  \label{fig3}
\end{figure}

Photoluminescence spectroscopy was performed on single Cu$_2$O microcrystals in a dilution refrigerator using continuous-wave green laser excitation (532\,nm), in particular analyzing the lineshape of the X$_O-\Gamma^-_{12}$ phonon-assisted orthoexciton transition. Three spectra acquired at laser input powers of 50\,nW, 5\,$\mu$W and 500\,$\mu$W are presented in Fig.\ref{fig3:sub1} together with fits using a Bose-Einstein distribution function in excellent agreement with the experimental lineshape. Extracted fit parameters for the chemical potential $\mu$ and the exciton gas temperature $T$ are shown in Fig.\ref{fig3:sub2} for excitation powers covering four orders of magnitude. Up to laser input powers of 10\,$\mu$W, the chemical potential was very close to zero ($\mu$=0 corresponds to the phase transition to a Bose-Einstein condensate), while the exciton gas temperature remained almost constant with values around 9\,K. The orthoexciton gas was gradually leaving the quantum-degenerate regime above excitation levels of 10\,$\mu$W, which coincides with deviations from the linear power dependence of orthoexciton luminescence (cf. Fig.\ref{fig2:sub4}). Additional data obtained from a different Cu$_2$O microcrystal with consistent orthoexciton gas characteristics is presented in Supplementary Fig.S6. Most interestingly, the extracted fit parameters suggest a different exciton gas behaviour compared to previous literature, where a quantum saturation effect was repeatedly observed with the exciton gas moving along adiabats parallel to the phase boundary of Bose-Einstein condensation \cite{Snoke1987,Lin1993,Snoke1990}. It was argued that the continuous temperature increase with increasing exciton density is associated with heating via two-body Auger decay. Here, the extracted fit parameters suggest increasing levels of quantum degeneracy and hence higher exciton densities for decreasing laser excitation at almost constant exciton gas temperatures. For natural bulk crystals, lineshapes suggesting quantum-degenerate statistics were attributed to a superposition of classical Maxwell-Boltzmann distributions as a result of exciton diffusion over several tens of micrometers \cite{OHara2000}. In our case, we can exclude this potential explanation due to the small sizes of the Cu$_2$O microcrystals. Further experimental and theoretical work will be required to assess if the presented orthoexciton gas characteristics are linked with Bose-Einstein condensation and to exclude possible alternative explanations, such as density-dependent recombination mechanisms or deviations from the ideal Bose gas theory. Our surprising results provide important new insights stimulating the use of Cu$_2$O microcrystals as experimental platform in future studies, which will be required to unambiguously solve the enigma of quantum-degenerate statistics and Bose-Einstein condensation of orthoexcitons in Cu$_2$O.

Furthermore, we demonstrate that paraexcitons are also in the quantum-degenerate regime by analyzing the lineshape of the phonon-assisted X$_P-\Gamma^-_{25}$ transition. Three spectra acquired under laser input powers of 10\,$\mu$W, 20\,$\mu$W and 40\,$\mu$W and the corresponding Bose-Einstein fits are shown in Fig.\ref{fig3:sub3}. We obtained high degrees of quantum degeneracy with fit values of the chemical potential $\mu$ around -0.1$\,k T$ and the temperature $T$ around 5\,K (10\,$\mu$W excitation). For increasing laser input powers, $\mu$ gradually decreased whereas $T$ increased. Note that deviations from the fit can be attributed to a phonon-assisted orthoexciton transition at higher energies (see also Fig.\ref{fig2:sub3}). Considering our results on orthoexcitons described above, Cu$_2$O microcrystals are capable of hosting quantum gas mixtures with intricate coupling between their components (spin-flip and spin-exchange processes \cite{Kavoulakis2000,Jang2004} of ortho- and paraexcitons) in confined geometries, providing new directions for future theoretical and experimental work. Moreover, the demonstration of quantum-degenerate paraexcitons in Cu$_2$O microcrystals render the latter a promising platform for exploring excitonic Bose-Einstein condensation in unprecedented device architectures, e.g. employing configurations for dynamic strain tuning by means of piezoelectric substrates \cite{Zeuner2018}. 

\floatsetup[figure]{style=plain,subcapbesideposition=top}

\subsection*{Rydberg excitons - the yellow \textit{\textbf{np}} series}
\begin{figure}[ht]
  \centering
  \sidesubfloat[]{\includegraphics[height=5.2cm]{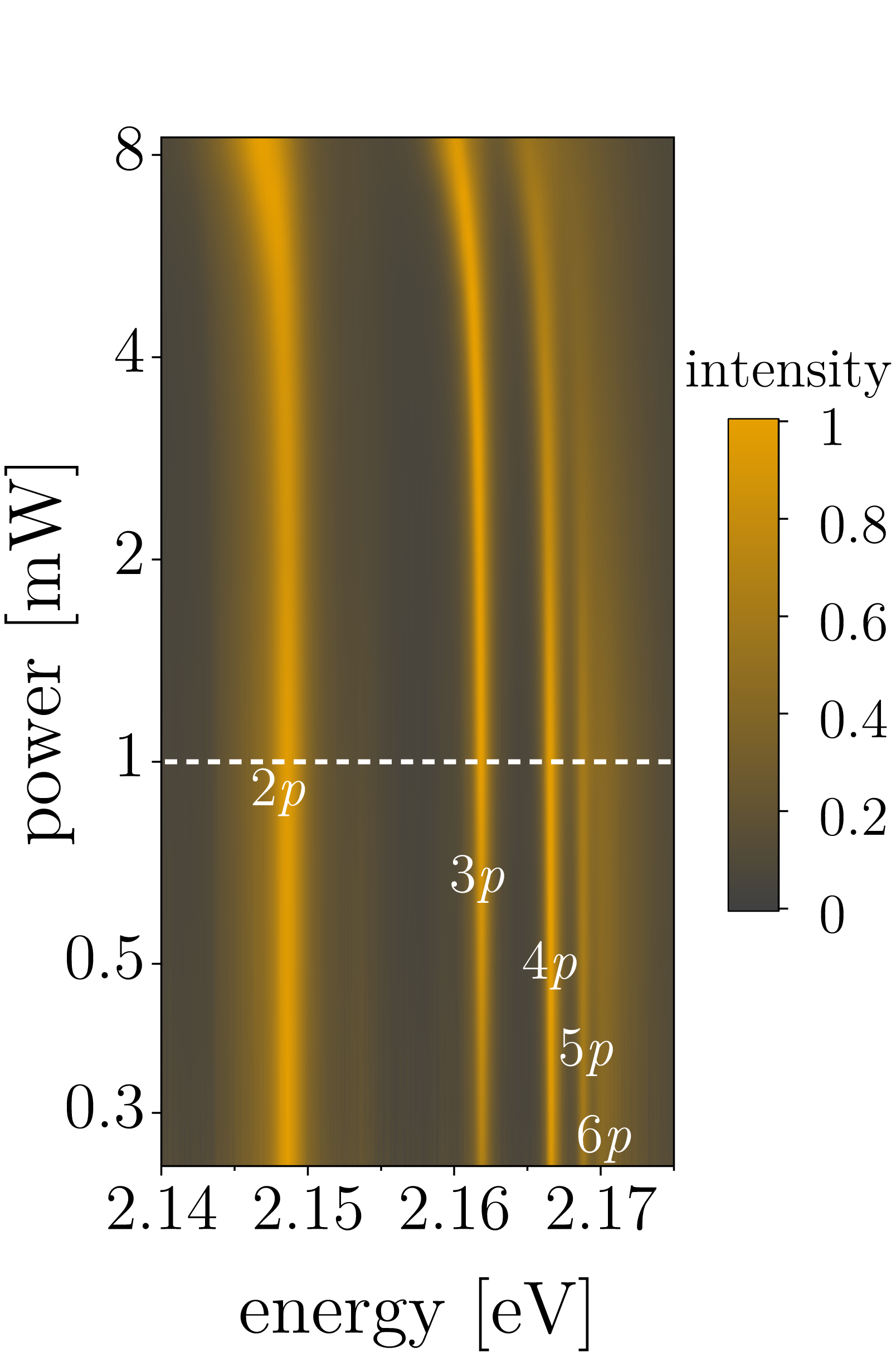}\label{fig4:sub1}}%
  \sidesubfloat[]{\includegraphics[height=5.2cm]{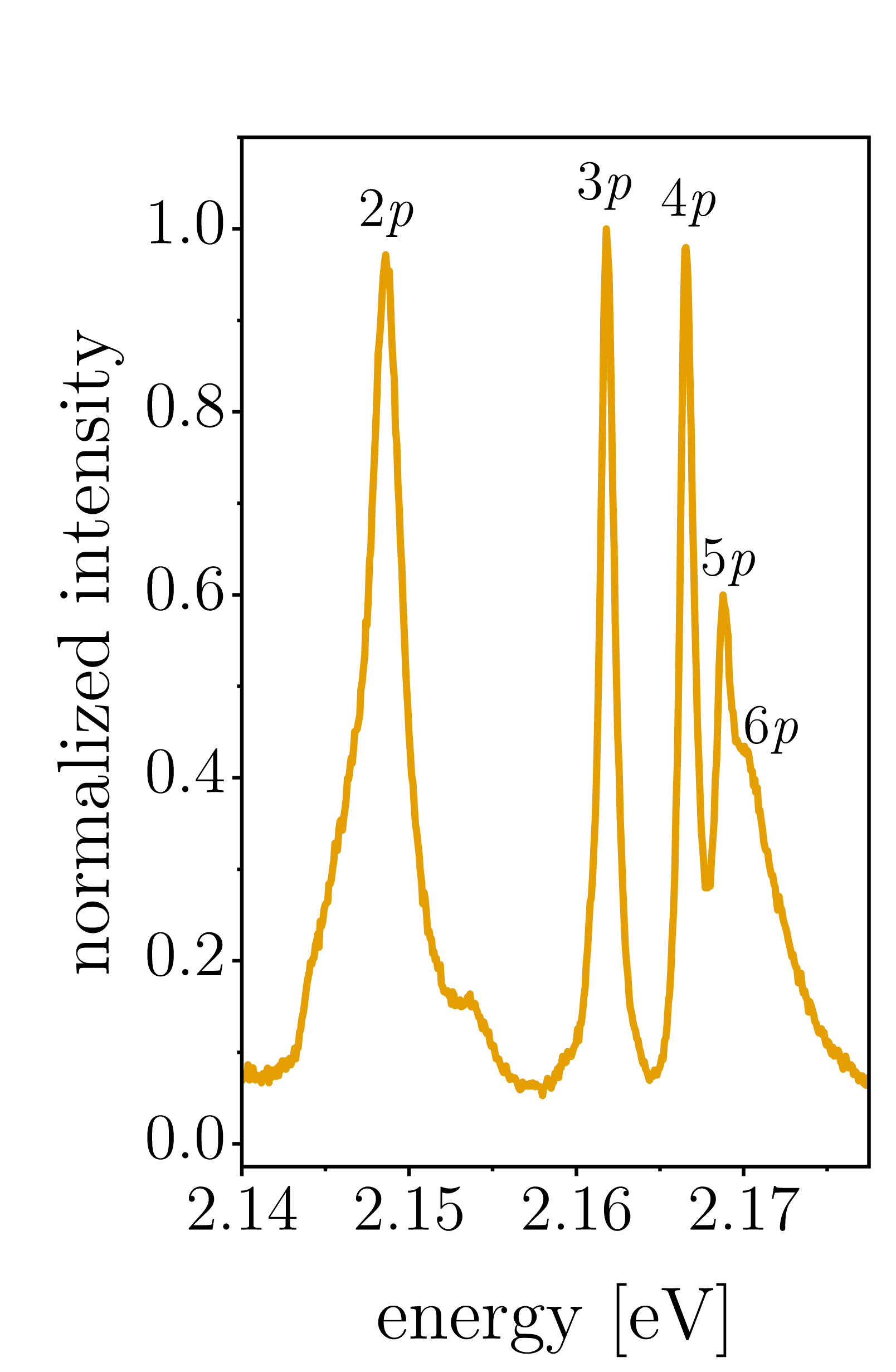}\label{fig4:sub2}}%
  \hspace{1.0mm}
  \sidesubfloat[]{\includegraphics[height=5.2cm]{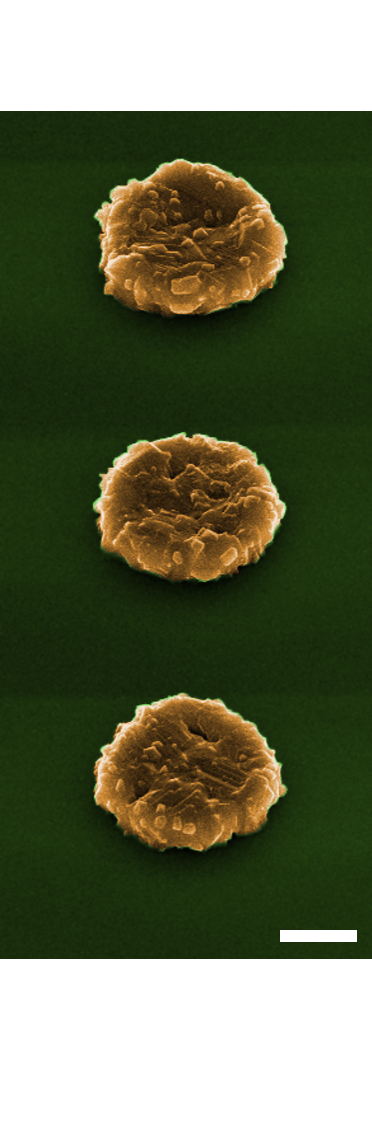}\label{fig4:sub3}}%
  \hspace{1.0mm}
  \sidesubfloat[]{\includegraphics[height=5.2cm]{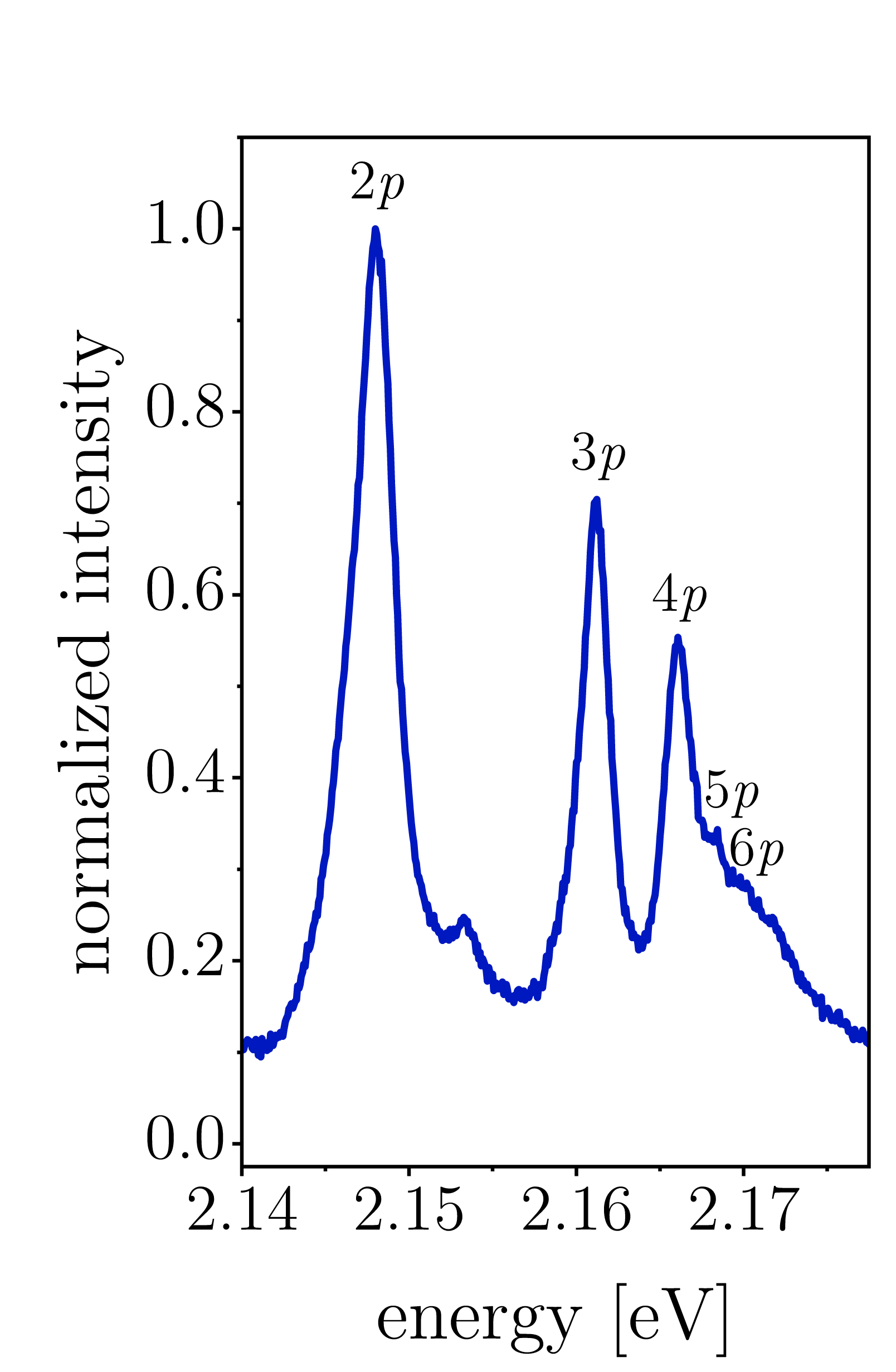}\label{fig4:sub4}}%
  \hspace{1.0mm}
  \sidesubfloat[]{\includegraphics[height=5.2cm]{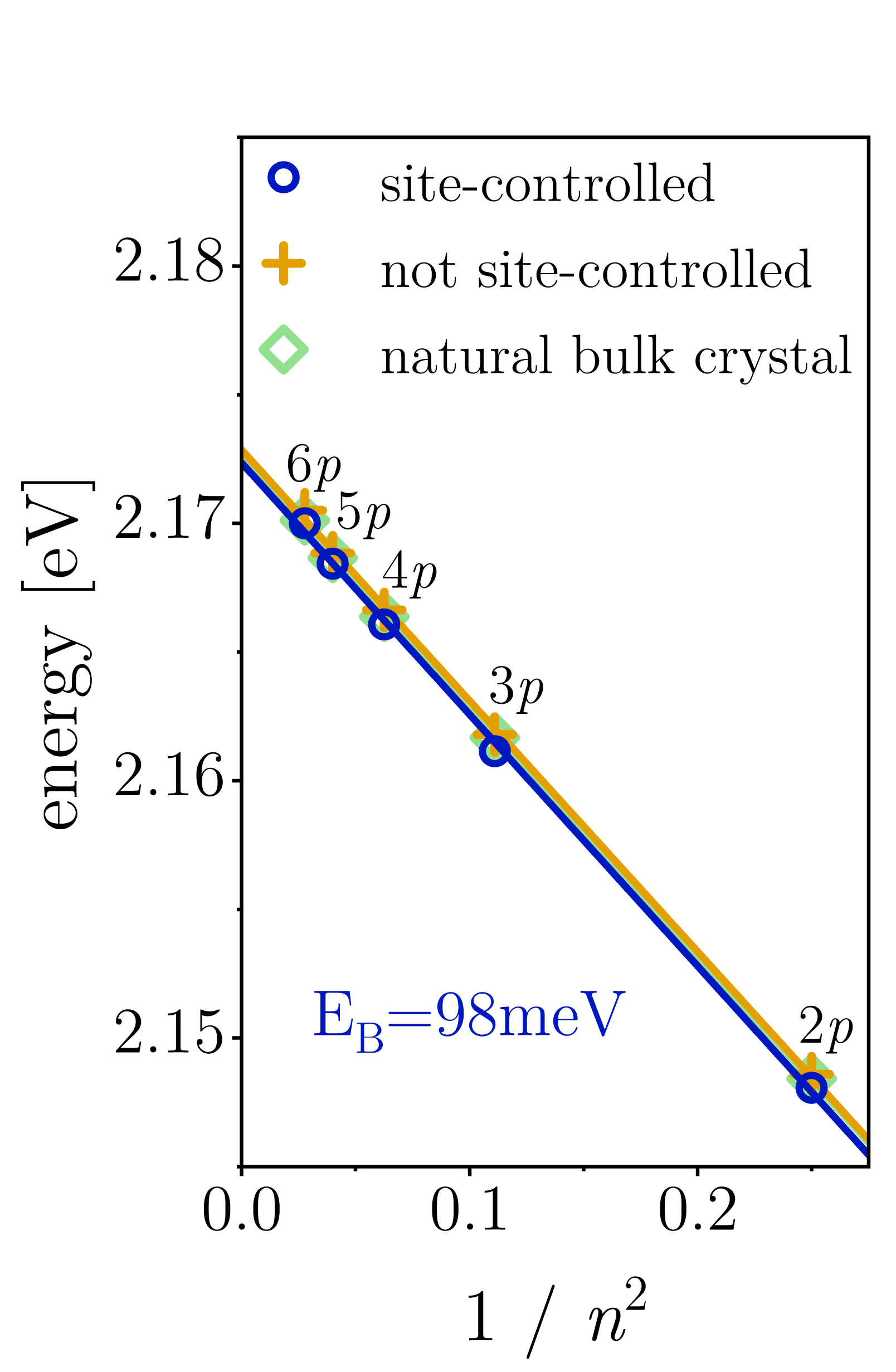}\label{fig4:sub5}}%
  \caption{\textbf{Fig.4 \textbar Rydberg excitons in Cu$_2$O microcrystals and site-controlled structures. a}, Normalized photoluminescence of \textit{np} Rydberg exciton emission from Cu$_2$O microcrystal for varying excitation powers. The white dashed line indicates the spectrum presented in \textbf{b}, which was acquired at an excitation power of 1\,mW and a cryostat stage temperature around 1.8\,K. \textbf{c}, Scanning electron micrograph of site-controlled circular Cu$_2$O microstructures with 5$\mu$m diameter (scale bar 2\,$\mu$m; sample tilt 45$^\circ$). \textbf{d} Photoluminescence spectrum of site-controlled Cu$_2$O microstructure that was acquired under identical experimental conditions as in b. \textbf{e}, Rydberg exciton energies as a function of $n^{-2}$ and the corresponding linear fits to extract the exciton binding energy E$_B$.}
  \label{fig4}
\end{figure}

After initial experiments in the middle of the last century, Rydberg excitons in Cu$_2$O have recently attracted considerable attention due to the experimental demonstration of principal quantum numbers up to \textit{n}=25 in absorption measurements using natural bulk crystals \cite{Kazimierczuk2014}. The question arises if Rydberg states can also be observed in the Cu$_2$O microcrystals presented here, which was assessed by means of photoluminescence experiments. Results obtained for varying incident laser powers are presented in Fig.\ref{fig4:sub1} (spectra normalized to their respective maxima). Emission peaks corresponding to Rydberg excitons up to \textit{n}=6 were identified at low and intermediate excitation, which can also be seen in the exemplary spectrum shown in Fig.\ref{fig4:sub2}. The high energy tail approaching the band gap could indicate the presence of higher-lying Rydberg excitons, which exhibit considerable spectral overlap due to their broadened luminescence linewidth. The 2\textit{p} peak showed a markedly asymmetric lineshape consistent with previous literature on bulk Cu$_2$O crystals \cite{Takahata2018,Kitamura2017,Reimann1989}, whereas the relative emission intensities from \textit{np} states differed in our case. The minor peak between the 2\textit{p} and 3\textit{p} energy level is attributed to \textit{s}-type excitons \cite{Takahata2018}. Rydberg exciton emission showed broadening for increasing excitation power, which can be explained by a combination of phonon scattering and density-dependent effects \cite{Kitamura2017}; the red shift indicates a bandgap decrease due to laser-induced sample heating. Most importantly, the power-dependent measurements verify the robustness of Rydberg excitons in Cu$_2$O microcrystals as their emission was detected in a wide range of excitation conditions. Additional photoluminescence spectroscopy data is shown in Supplementary Fig.S7, validating that Rydberg excitons could be consistently observed in Cu$_2$O microcrystal samples.

Furthermore, site-controlled Cu$_2$O microstructures (Fig.\ref{fig4:sub3}) were achieved by lithographic patterning of the copper film before oxidation. We demonstrate luminescence from excited Rydberg states in a circular Cu$_2$O microstructure with 5$\mu$m diameter (Fig.\ref{fig4:sub4}). The intensity ratio of \textit{np} states was different compared to Cu$_2$O microcrystals, which could be explained by variations in microscopic strain \cite{Reimann1989}. The Rydberg exciton energies were evaluated as a function of $n^{-2}$ for results obtained from site-controlled microstructures, from not site-controlled microcrystals and from a natural bulk crystal (Fig.\ref{fig4:sub5}). Excellent agreement with a linear relation was found and exciton binding energies of 98\,meV and 97\,meV were deduced for the synthetic samples and the natural bulk crystal, respectively. The extracted exciton binding energies concur with previous findings using bulk crystals, obtained from both photoluminescence (97\,meV\cite{Kitamura2017}; 98.5\,meV\cite{Reimann1989}) and absorption measurements (98\,meV\cite{Matsumoto1996}). Hence our results demonstrate the realization of site-controlled, micrometer-sized Cu$_2$O structures as host platform for Rydberg excitons on silicon, opening up opportunities for unprecedented photonic device architectures. For instance, our technology will enable applications in nonlinear quantum optics relying on interactions between Rydberg states \cite{Firstenberg2016}, as clear signatures of the Rydberg blockade effect have been recently reported for principal quantum numbers around \textit{n}=6 \cite{Heckotter2018}. We expect that in the presented synthetic Cu$_2$O samples the observation of Rydberg states with higher principal quantum number \textit{n} is not limited by material quality but is hampered by luminescence broadening, similar to previous reports on natural crystals \cite{Kitamura2017}, suggesting future experiments in an absorption geometry. 

\floatsetup[figure]{style=plain,subcapbesideposition=top}

\subsection*{Conclusion}
We have presented the growth of Cu$_2$O microcrystals on silicon substrates, showing excellent optical material quality with exceedingly low point defect and impurity levels. The fabrication method for obtaining high-quality Cu$_2$O films via a scalable thermal oxidation process has guiding significance for diverse areas where this low-cost, non-toxic material is employed, such as photovoltaics and photocatalysis. Cu$_2$O microcrystals were identified as ideal host material for dense exciton gases at milli-Kelvin temperatures, in particular 1\textit{\textbf{s}} ortho- and paraexcitons exhibiting kinetic energy distributions obeying Bose-Einstein statistics. Excitons in the quantum-degenerate regime were obtained through continuous-wave laser excitation of confined micrometer-sized geometries, which constitutes an entirely new approach for assessing the feasibility of excitonic Bose-Einstein condensation in this material. Furthermore, the demonstration of Rydberg excitons in Cu$_2$O microcrystals and their integration on silicon have far-reaching implications for future applications in photonic quantum information processing. For instance, Rydberg states in Cu$_2$O have been proposed for the realization of novel solid-state single-photon sources \cite{Khazali2017} and giant optical nonlinearities \cite{Walther2018}. Hence our work lays the foundation for a platform technology enabling solid-state Rydberg excitations on-chip, which is envisioned to result in integrated devices capable of generating and manipulating light at the single photon level.

\small
\section*{Methods}
\subsection*{Growth of Cu$_2$O microcrystals on silicon substrates}
The deposition of copper films (thickness $\sim$700\,nm) was performed by electron beam evaporation onto pieces of silicon wafers with 150\,nm thermal SiO$_2$. A thin intermediate titanium layer (thickness $\sim$5\,nm) was employed to improve the film adhesion on the substrate surface. Samples with structured Cu$_2$O were realized by patterning the copper film using an electron beam lithography lift-off process. Thermal oxidation was carried out in a tube furnace connected to a vacuum pump. Before growth experiments the system was evacuated and purged multiple times using high-purity synthetic air (Air Liquide Alphagaz 2). Cu$_2$O samples were grown by thermal oxidation at a a pressure around 1\,mbar and setpoint temperatures of 800$^\circ$C or 850$^\circ$C. The temperature was kept constant for 1\,h or 5\,h after reaching the setpoint value, followed by natural sample cool-down. 

\subsection*{Sample characterization}
The morphologies of Cu$_2$O films and microstructures were characterized by scanning electron microscopy imaging of the sample surfaces and of cross-sections obtained by mechanical cleaving. X-ray diffraction measurements were performed by specular scans using a PANalytical Empyrean system. Radiation from a sealed copper tube was used in combination with a secondary graphite monochromator in Bragg–Brentano geometry. Phase analysis was carried out relying on powder patterns from the database PDF2, International Center for Diffraction Data, using 004-0836 for Cu and 005-0667 for Cu$_2$O. 

\subsection*{Photoluminescence spectroscopy and data evaluation}
All photoluminescence spectroscopy experiments were performed using a Horiba iHR550 spectrometer and a continuous-wave, diode-pumped solid-state laser emitting at 532\,nm. Room-temperature spectra were acquired using an objective (NA=0.82) for excitation and collection. Spectroscopy at milli-Kelvin temperatures was performed relying on a cryogen-free dilution refrigerator (Bluefors) with a base temperature around 10\,mK. Optical side-access windows with anti-reflective coatings were used for laser excitation of the samples mounted on a dedicated stage controlled by piezoelectric actuators, which had a base temperature around 40\,mK. The laser was focused by a lens (NA=0.50) inside the cryostat to a spot diameter around 1.2\,$\mu$m (full width at half maximum); the power was measured at the outermost cryostat window. Spectroscopy results on phonon-assisted ortho- and paraexciton emission at milli-Kelvin temperatures were fitted using a Bose-Einstein distribution function. Convolution with a Gaussian function (orthoexcitons $\sigma$=0.20\,meV; paraexcitons $\sigma$=0.12\,meV) was considered to account for additional broadening by the linewidth of the direct orthoexciton transition and by the spectrometer resolution.

\section*{Acknowledgements}
The authors thank Roland Resel for helpful discussions. The Quantum Nano Photonics Group at KTH acknowledges financial support from the Linnaeus Center in Advanced Optics and Photonics (ADOPT). M.A.M.V. acknowledges funding from the Swedish Research Council under grant agreement No. 2016-04527. V.Z. acknowledges funding by the European Research Council under grant agreement No. 307687 (NaQuOp) and the Swedish Research Council under grant agreement \mbox{No. 638-2013-7152}. The project was co-funded by Vinnova and FP7 (GROWTH 291795).

\section*{Author contributions statement}
S.S., M.A.M.V. and V.Z. conceived the experiments, with input from A.M. M.A.M.V. designed and built the milli-Kelvin photoluminescence spectroscopy setup. S.S. performed material growth, SEM characterization and photoluminescence experiments. Data analysis and interpretation was performed by S.S. with support from M.A.M.V., A.M. and V.Z. XRD characterization and the corresponding analysis was performed by B.K. The project was supervised by M.A.M.V. and V.Z. The manuscript was written by S.S. with inputs from all authors.

\section*{Additional information}
The authors declare no competing financial interests.

%\begin{document}
\flushbottom
%\maketitle
\newpage
\section*{Supplementary Figures}

\vspace{10mm}
\floatsetup[figure]{style=plain,subcapbesideposition=top}
\begin{figure}[h]
  \centering
  \includegraphics[width=0.45\textwidth]{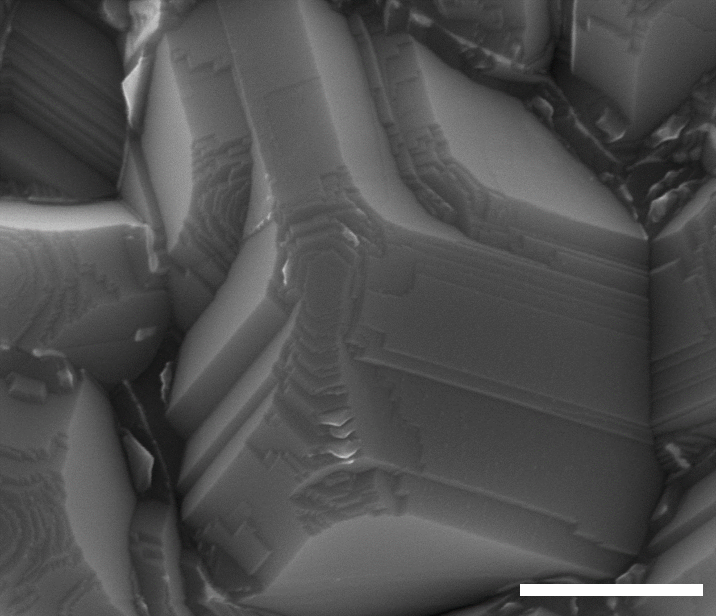}\label{figS1:sub1}
  \caption*{\textbf{Fig.S1 \textbar High-resolution scanning electron microscope image of Cu$_2$O microcrystal surface.} Extended terrace-like structures suggest a two-dimensional growth mode for the individual microcrystals (scale bar 500\,nm).}
  \label{figS1}
\end{figure}

\vspace{32mm}
\floatsetup[figure]{style=plain,subcapbesideposition=top}
\begin{figure}[h]
  \centering
  \sidesubfloat[]{\includegraphics[width=0.45\textwidth]{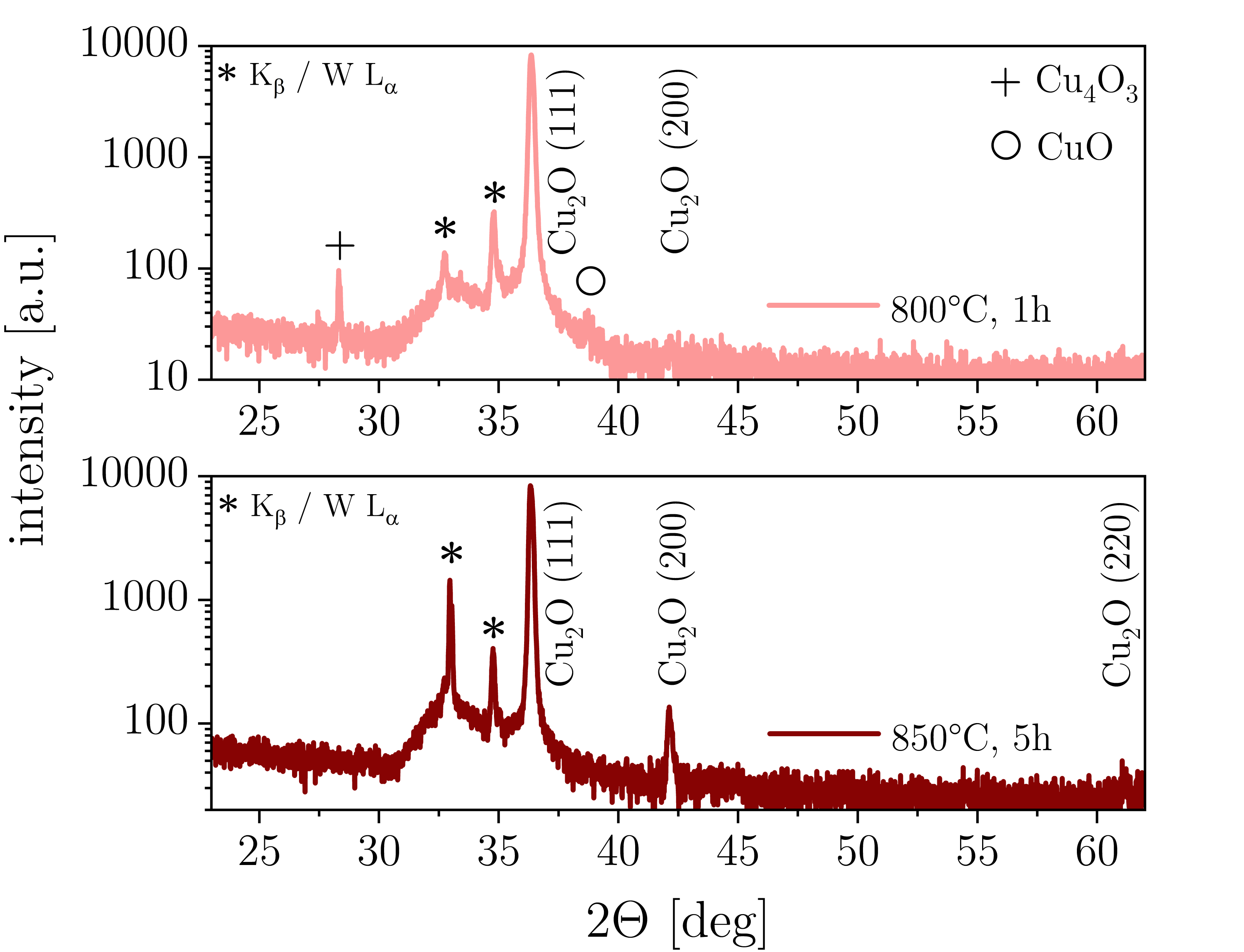}\label{figS2:sub1}}%
  \sidesubfloat[]{\includegraphics[width=0.48\textwidth]{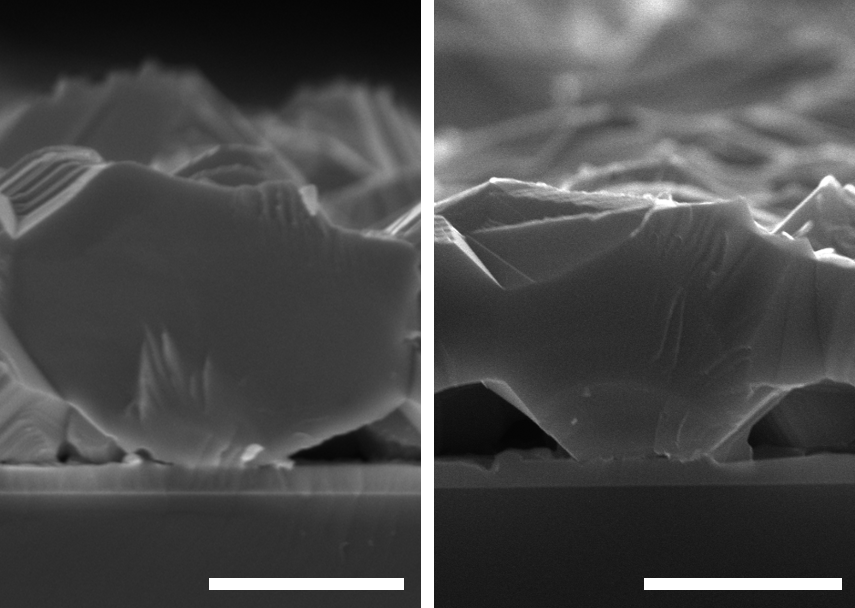}\label{figS2:sub2}}%
  \caption*{\textbf{Fig.S2 \textbar Characterization of Cu$_2$O microcrystal samples grown at different conditions. a}, X-ray diffraction of films obtained through thermal oxidation at 800$^\circ$C for 1\,h and at 850$^\circ$C for 5\,h at pressures around 1\,mbar of synthetic air. The former sample showed residual Cu$_4$O$_3$ and CuO phases, whereas in the latter sample phase-pure Cu$_2$O was found (database PDF2, International Center for Diffraction Data, using 005-0667 for Cu$_2$O, 049-1830 for Cu$_4$O$_3$ and 045-0937 for CuO). \textbf{b}, The corresponding cross-sectional scanning electron microscope images (left: 800$^\circ$C / 1\,h, right: 850$^\circ$C / 5\,h) show comparable Cu$_2$O microcrystal morphology (scale bars 1\,$\mu$m). For the longer growth time a higher degree of intergrain connectivity was observed.}
  \label{figS2}
\end{figure}

\clearpage
\vspace*{21mm}
\floatsetup[figure]{style=plain,subcapbesideposition=top}
\begin{figure}[h]
  \centering
  \includegraphics[width=0.45\textwidth]{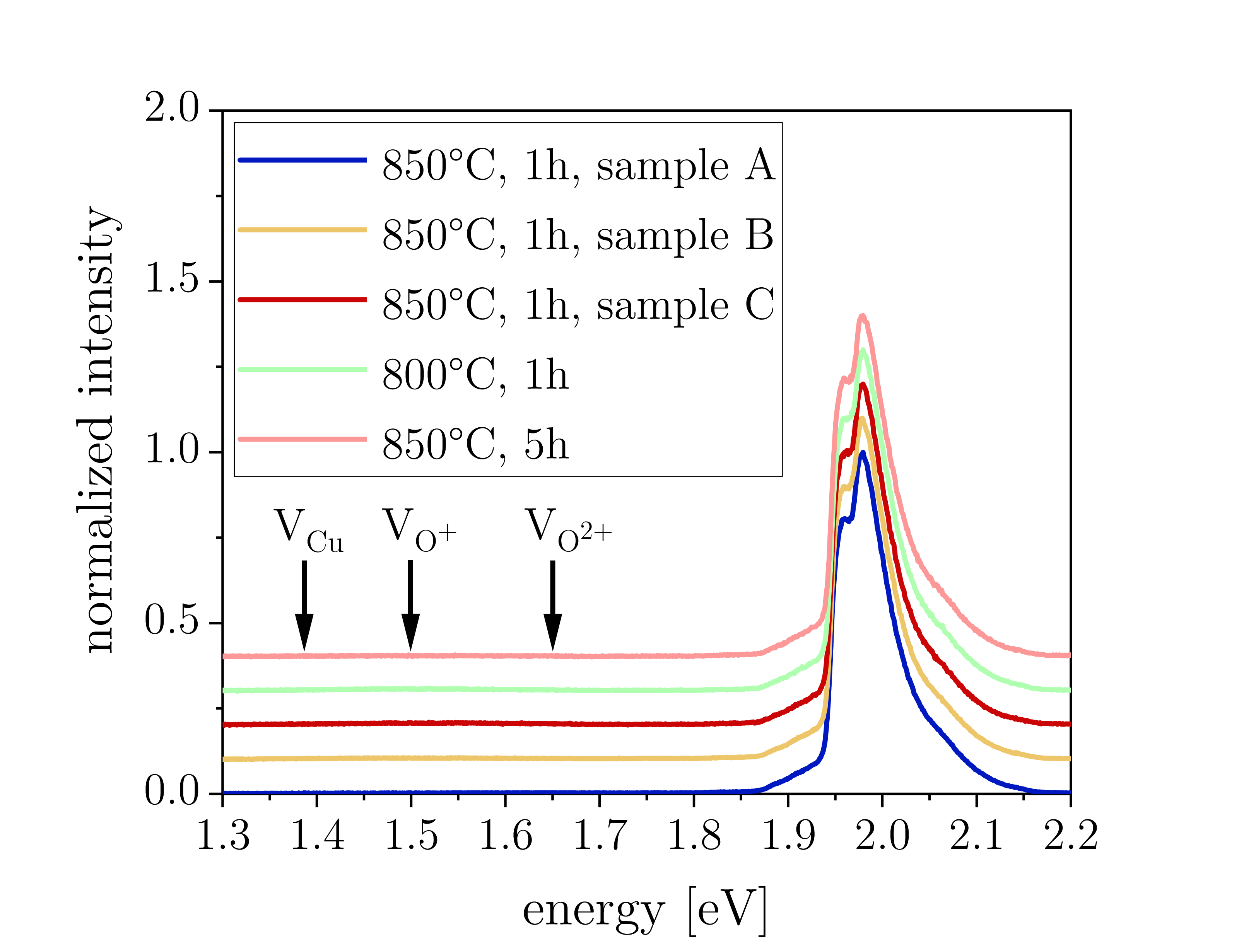}
  \caption*{\textbf{Fig.S3 \textbar Room-temperature photoluminescence spectroscopy of excitons and point defects.} The comparison of three samples obtained from different batches (850$^\circ$C / 1\,h) and two samples grown under different conditions (800$^\circ$C / 1\,h; 850$^\circ$C / 5\,h) show very similar excitonic emissions and no marked luminescence from point defects in all cases (spectra were offset vertically for clarity).}
  \label{figS3}
\end{figure}

\vspace{28.5mm}
\floatsetup[figure]{style=plain,subcapbesideposition=top}
\begin{figure}[h]
  \centering
  \sidesubfloat[]{\includegraphics[width=0.45\textwidth]{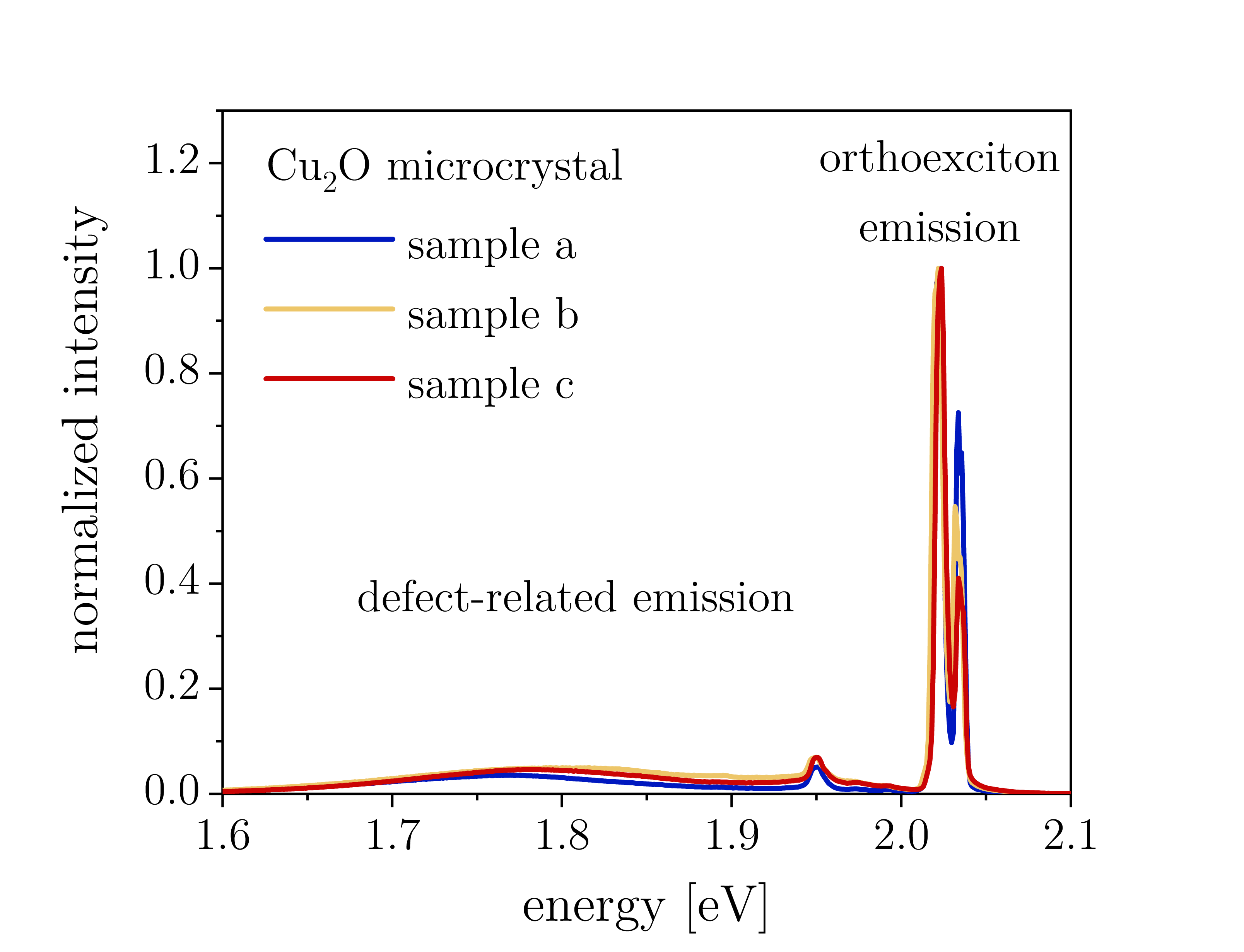}\label{figS4:sub1}}%
  \sidesubfloat[]{\includegraphics[width=0.45\textwidth]{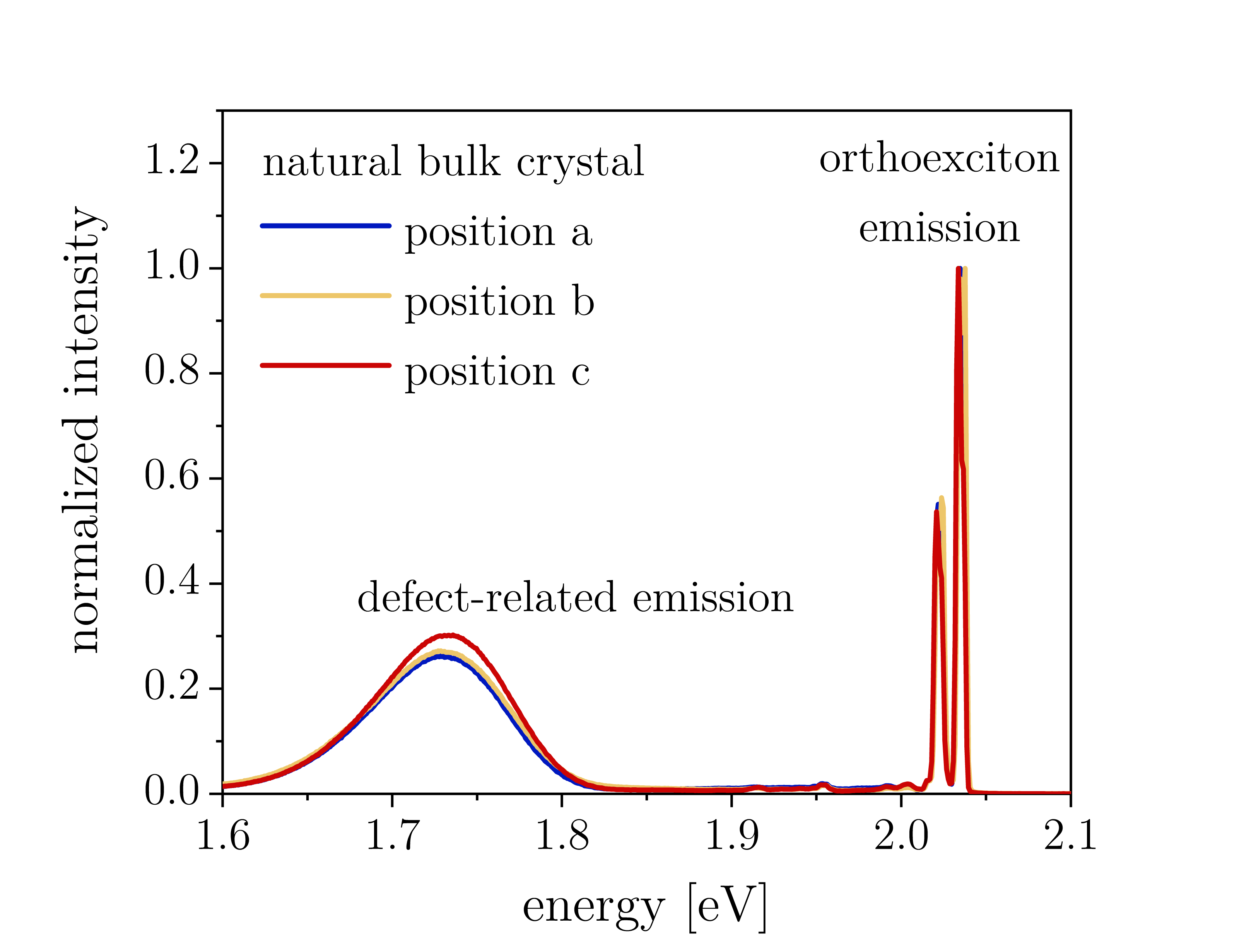}\label{figS4:sub2}}%
  \caption*{\textbf{Fig.S4 \textbar Photoluminescence spectroscopy of excitons and point defects at milli-Kelvin temperatures. a}, Comparison of emission from Cu$_2$O microcrystals on three different samples (850$^\circ$C / 1\,h), which all show exceedingly low point defect levels. \textbf{b}, Comparison of three different sample positions on the natural bulk crystal, exhibiting comparable emission from oxygen vacancies. All measurements were performed at an excitation power of 50\,$\mu$W.}
  \label{figS4}
\end{figure}

\clearpage
\vspace*{21mm}
\floatsetup[figure]{style=plain,subcapbesideposition=top}
\begin{figure}[h]
  \centering
  \includegraphics[width=0.45\textwidth]{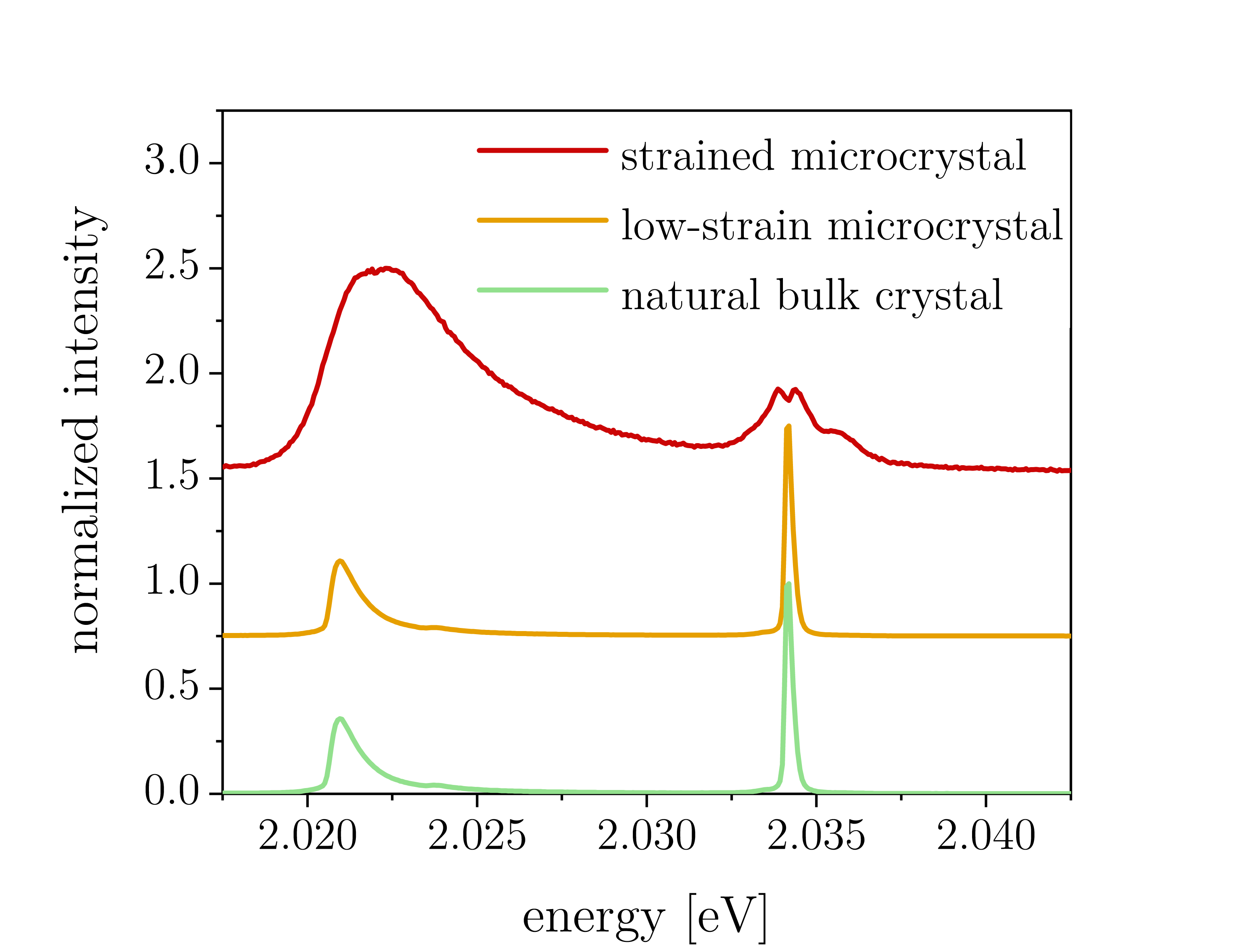}
  \caption*{\textbf{Fig.S5 \textbar The influence of microscopic strain on excitons in Cu$_2$O microcrystals.} Comparison of phonon-assisted and direct orthoexciton emission for strained Cu$_2$O microcrystal, low-strain Cu$_2$O microcrystal and the natural bulk crystal (excitation power 50\,$\mu$W; spectra vertically offset for clarity). Strained Cu$_2$O microcrystals exhibited a lifted energy degeneracy of the triplet orthoexciton state, similar to strained bulk crystals \cite{Lin1993}, with energy splittings around 1.5\,meV, which is considerably smaller compared to previous reports on Cu$_2$O thin films epitaxially grown on MgO substrates \cite{Sun2002,Aihara2015}. Stress values around 50\,MPa were estimated from the experimental data assuming rhombohedral stress along the [110] direction \cite{Trebin1981}.}
  \label{figS5}
\end{figure}

\vspace*{21.5mm}
\floatsetup[figure]{style=plain,subcapbesideposition=top}
\begin{figure}[h]
  \centering
  \sidesubfloat[]{\includegraphics[width=0.3\textwidth]{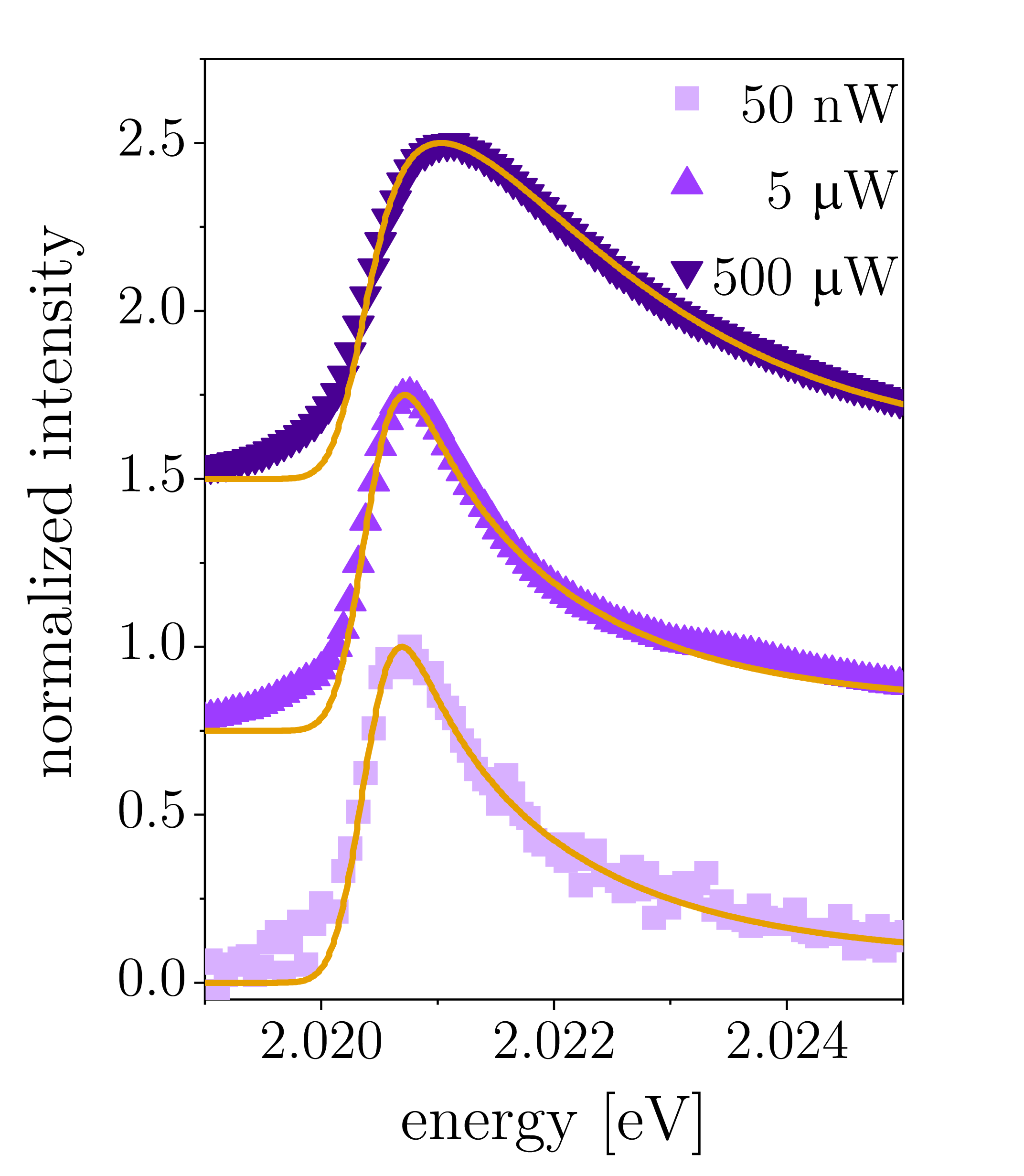}\label{figS6:sub1}}%
  \sidesubfloat[]{\includegraphics[width=0.3\textwidth]{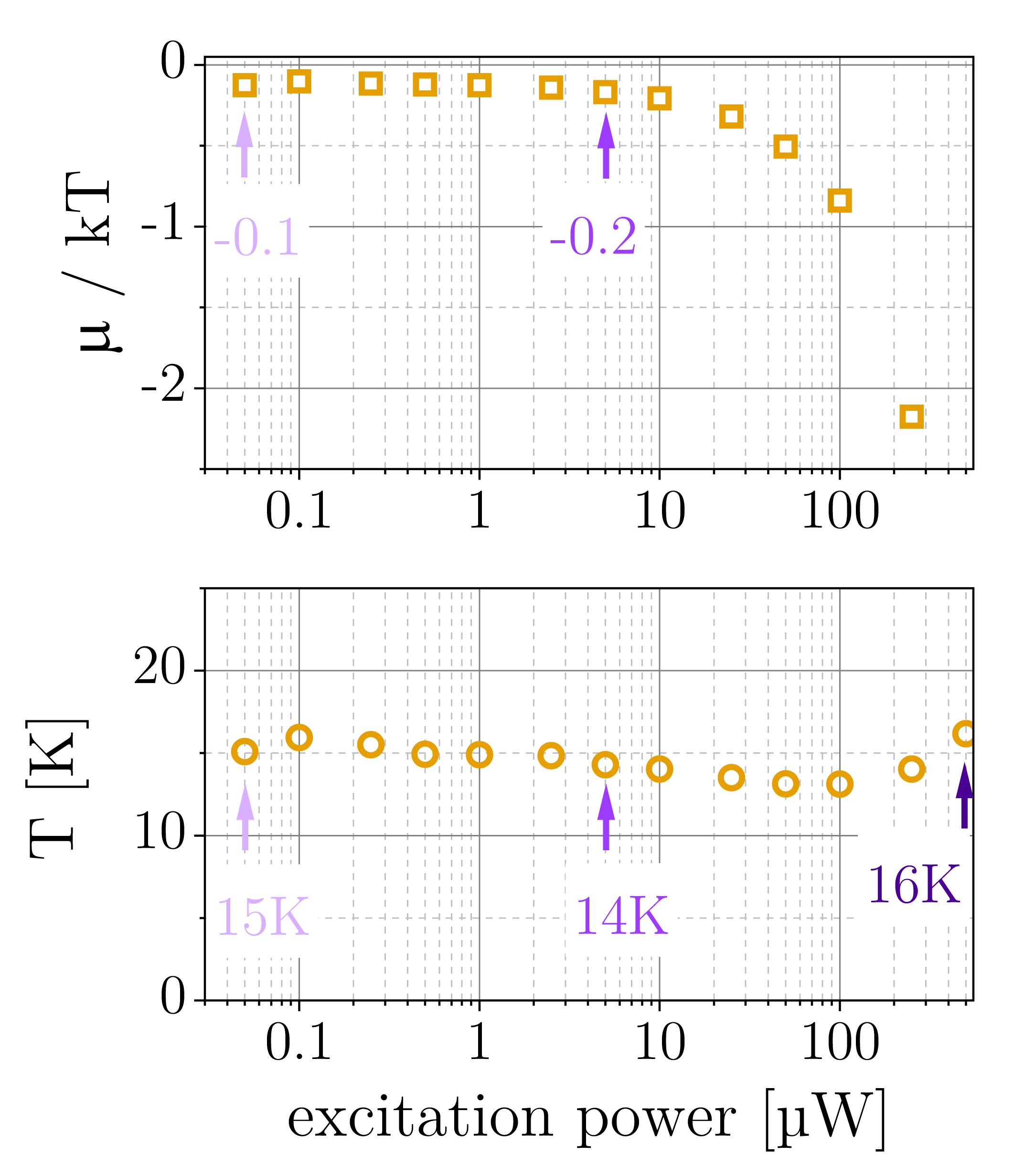}\label{figS6:sub2}}%
  \caption*{\textbf{Fig.S6 \textbar Quantum-degenerate orthoexciton gas in another Cu$_2$O microcrystal.} Additional measurements were performed on a different sample grown under identical conditions. \textbf{a}, Photoluminescence spectra of phonon-assisted X$_O-\Gamma^-_{12}$ orthoexciton transition for three different excitation powers. The solid lines correspond to fits using a Bose-Einstein distribution function (spectra were offset vertically for clarity). \textbf{b}, Extracted fit parameters for the chemical potential $\mu$ and the temperature \textit{T} of the orthoexciton gas for excitation powers covering four orders of magnitude. The parameters for the spectra shown in a, are annotated in the graph (The fit obtained for 500\,$\mu$W excitation resulted in Maxwell-Boltzmann-like statistics and hence the value for the chemical potential is outside the plot range).}
  \label{figS6}
\end{figure}

\clearpage
\vspace*{25mm}
\floatsetup[figure]{style=plain,subcapbesideposition=top}
\begin{figure}[h]
  \centering
  \includegraphics[height=5.7cm]{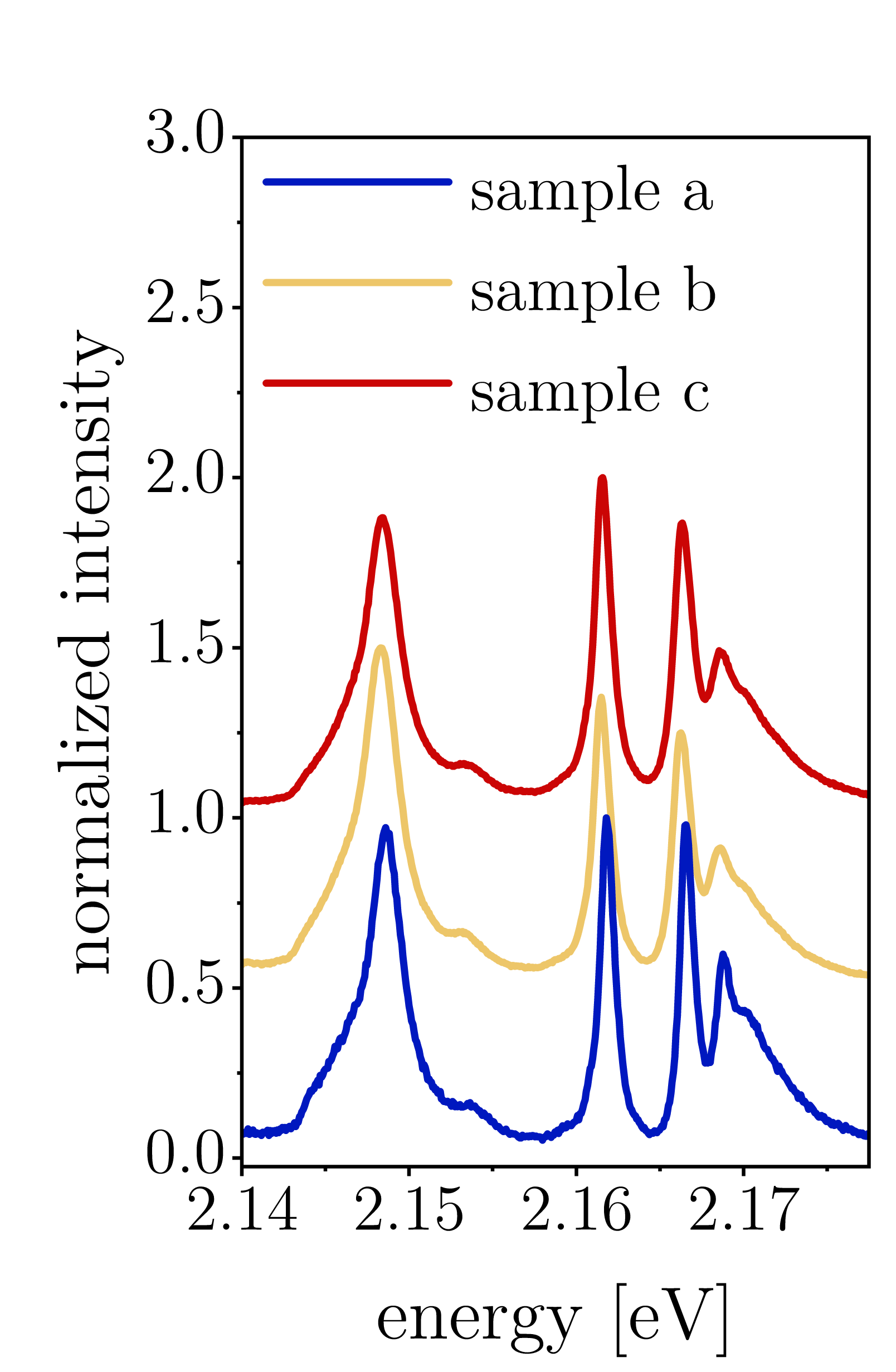}
  \caption*{\textbf{Fig.S7 \textbar Luminescence from Rydberg excitons in Cu$_2$O microcrystals.} Comparison of results obtained from three different samples (850$^\circ$C / 1\,h), exhibiting very similar characteristics (excitation power 1\,mW; spectra vertically offset for clarity).}
  \label{figS7}
\end{figure}

%\end{document}
\end{document}